\documentclass[twocolumn]{aastex62}
\usepackage[varg]{txfonts}
\usepackage{graphicx}
\usepackage{color}
\usepackage{eqnarray}
\begin{document}

\title{Kinematical signs of dust trapping and feedback in a local pressure bump in the protoplanetary disk around HD~142527 revealed with ALMA}

\author{Hsi-Wei Yen}
\affiliation{Academia Sinica Institute of Astronomy and Astrophysics, 11F of Astro-Math Bldg, 1, Sec. 4, Roosevelt Rd, Taipei 10617, Taiwan}
\author{Pin-Gao Gu}
\affiliation{Academia Sinica Institute of Astronomy and Astrophysics, 11F of Astro-Math Bldg, 1, Sec. 4, Roosevelt Rd, Taipei 10617, Taiwan}

\correspondingauthor{Hsi-Wei Yen}
\email{hwyen@asiaa.sinica.edu.tw}

\begin{abstract}
We analyzed the archival data of the continuum emission at six wavelengths from 3 to 0.4 mm and $^{13}$CO and C$^{18}$O (1--0, 2--1, and 3--2) lines in the protoplanetary disk around HD~142527 obtained with the Atacama Large Millimeter/submillimeter Array. We performed fitting to the spectral energy distributions obtained at the six wavelengths with the gray-body slab models to estimate the distributions of the dust surface density and spectral index of dust absorption coefficient $\beta$. We also estimated the distribution of the gas column density by fitting the C$^{18}$O spectra and measured the disk rotation by fitting the Keplerian disk models to the C$^{18}$O data. We found super- and sub-Keplerian rotation inside and outside the dust ring in the northwest in the HD~142527 disk, suggestive of the presence of a local pressure bump. In comparison with our estimated dust and gas distributions, the location of the pressure bump is coincident with the region showing a three times higher dust density and a three times lower gas-to-dust mass ratio than the mean values in the disk, suggesting dust trapping in the pressure bump. Nevertheless, there is no correlation between our derived $\beta$ distribution and the location of the pressure bump.
In addition, we found that the width of the dust ring is comparable or larger than the width of the pressure bump, which could suggest that dust feedback is significant in the pressure bump.
\end{abstract}


\section{Introduction}
Growth of dust grains to large aggregates and further to planetesimals is an important step in the standard core accretion scenario of planet formation \citep{Helled14,Raymond14}.
Thus, observational studies on the mechanisms of dust concentration, which could lower gas-to-dust mass ratio and may eventually enable grain growth to planetesimals \citep{Chiang10,Testi14}, in protoplanetary disks are essential to understand the planet formation processes. 
In a protoplanetary disk, 
rotational velocity of gas is determined by the balance between gravity, centrifugal force, and pressure gradient, 
while dust grains, which are not affected by the pressure, tend to orbit around the center in Keplerian rotation. 
The relative motion between gas and dust results in drag force on dust grains and makes dust grains to drift toward a pressure maximum in a protoplanetary disk, leading to dust trapping and concentration \citep{Birnstiel10,Chiang10,Testi14}. 
This process also depends on dust properties and gas density.  
Therefore, comparing the dust distributions with the deviation of gas motion from Keplerian rotation in protoplanetary disks could shed light on dust properties and mechanisms of dust concentration \citep[e.g.,][]{Perez18, Teague18, Zhang18, Rosotti20}. 

Transitional disks, which have an inner cavity and dust rings observed at millimeter wavelengths \citep{Williams11}, are promising targets to study dust trapping and concentration, where dust grains can drift toward and be trapped at the pressure bump at the edge of the inner cavity \citep{Pinilla12}. 
This has been proposed to explain the gas and dust distributions in several transitional disks \citep[e.g.,][]{Marel16b, Fedele17, Pinilla17}.
In addition, a group of transitional disks exhibit asymmetric, crescent features in the millimeter continuum emission \citep[e.g.,][]{Perez14, Casassus15, Marel16}, suggestive of enhancement in the dust density in certain azimuthal directions. 
Azimuthally asymmetric dust concentration could be due to dust trapping at the pressure maximum of a vortex formed at the edge of the inner cavity \citep{Lin14,Zhu14,Baruteau16}, eccentric disk rotation excited by a massive planet or companion \citep{Hsieh12,Ataiee13,Ragusa17}, or asymmetric spirals driven by a massive companion \citep{Price18}. 
These different mechanisms are expected to exhibit different features in the gas kinematics \citep[e.g.][]{Hsieh12,Huang18,Price18,Pinte19}.
Thus, comparison between the dust distributions and the gas kinematics in protoplanetary disks could also reveal the origins of the dust concentration. 

The protoplanetary disk around HD~142527 is one of the disks showing a high-contrast crescent dust ring at millimeter wavelengths \citep{Ohashi08,Casassus13}.
HD~142527 is a binary system of two pre-main sequence stars with a mass ratio of 0.1--0.2  \citep{Biller12, Lacour16, Christiaens18}. 
The distance to HD~142527 is 156 pc, measured with the {\it Gaia} mission \citep{Gaia16,Gaia18}. 
The HD~142527 disk has an inner cavity with a radius of $\sim$100 au \citep{Ohashi08,Muto15,Perez15,Boehler17}.
Several spiral arms driven by the companion have been observed in infrared and the CO lines in the HD~142527 disk \citep{Fukagawa06, Casassus12, Avenhaus14, Avenhaus17, Christiaens14, Rodigas14}.
Observations of the thermal dust emission at centimeter wavelengths and the polarized millimeter continuum emission due to dust scattering suggest the presence of large dust grains with sizes larger than 150 $\mu$m in the crescent dust ring \citep{Casassus15,Kataoka16,Ohashi18}.
With the observations of the Atacama Large Millimeter/submillimeter Array (ALMA) in one or two frequency bands, 
the gas and dust distributions in the HD~142527 disk were analyzed \citep{Muto15, Boehler17, Soon19}. 
These ALMA results suggest a low gas-to-dust mass ratio in the crescent dust ring and hint at dust trapping, 
but the relations between the gas kinematics and the crescent dust ring have not yet been studied in detail. 
Therefore, the HD~142527 disk is an excellent target to observationally study the mechanisms of dust concentration. 

HD~142527 has been observed with ALMA in six different frequency bands to date. 
Thus, in this work, we revisit the questions of the gas and dust distributions in the HD~142527 disk with joint analysis of the continuum emission at six different wavelengths from 3~mm to 0.4~mm and the CO isotopologue lines at multiple transitions to minimize the uncertainties due to the continuum opacity and the excitation temperature of the lines.
In addition, we analyze the gas kinematics traced by the CO isotopologue lines in comparison with Keplerian rotation, 
and compare the observed non-Keplerian rotation with the dust and gas distributions to study the dust concentration and trapping in the HD~142527 disk. 
The present paper is organized as follows. 
In section 2, we introduce the data included in our analysis. 
In section 3, we describe our analysis on the gas and dust distributions and the gas kinematics in the HD~142527 disk. 
In section 4, we discuss the signs of a pressure bump in the disk revealed from the gas kinematics and its relations with the gas and dust distributions. 
Finally, through the comparison between the gas kinematics and the gas and dust distributions, we discussed our results in the context of dust trapping and feedback in a pressure bump.

\section{Data}
We retrieved the data of the continuum emission at 3~mm, 2~mm, 1.3~mm, 0.9~mm, 0.6~mm, and 0.4~mm and the $^{13}$CO and C$^{18}$O emission lines of $J=1\mbox{--}0$, 2--1, and 3--2 from the ALMA archive. 
The data were obtained with the ALMA projects of 
2012.1.00631.S, 2012.1.00725.S, 2013.1.00305.S, 2013.1.00670.S, 2015.1.00614.S, 2015.1.01137.S, 2015.1.01353.S, and 2017.1.00987.S. 
The basic parameters of these observations are summarized in Table \ref{obs}.
We calibrated the raw visibility data obtained from the archive using the software Common Astronomy Software Applications \citep[CASA;][]{McMullin07} with the calibration scripts or pipelines provided by the ALMA observatory. 
We additionally performed self calibration of phase on the continuum data, 
and the continuum was subtracted from the molecular-line data in the {\it uv} domain.

\begin{deluxetable}{ccccc}
\tablecaption{Summary of observations}
\centering
\tablehead{Project & Antenna & Frequency & Baseline & On source \\
code & number & band & length & time}
\startdata
2012.1.00631.S & 36--39 & 7 &  17--1789 k$\lambda$ & 114 mins\\
2012.1.00725.S & 37 & 7 & 24--891 k$\lambda$ & 63 mins\\
2013.1.00305.S & 43 & 6 & 27--1211 k$\lambda$ & 34 mins\\
2013.1.00670.S & 37--40 & 3 & 9--525 k$\lambda$ & 226 mins\\
2015.1.00614.S & 41 & 9 & 36--1095 k$\lambda$ & 23 mins\\ 
2015.1.01137.S & 40--42 & 8 & 23--1016 k$\lambda$ & 24 mins\\
2015.1.01353.S & 38--40 & 6 & 12--865 k$\lambda$ & 61 mins\\
2017.1.00987.S & 43 & 4 & 7--1216 k$\lambda$ & 116 mins 
\enddata
\end{deluxetable}\label{obs}

The calibrated visibility data of the same lines or the continuum in the same frequency bands were combined to generate images. 
All the images were generated with the Briggs weighting. 
The adopted robust parameters and other basic parameters of the images are summarized in Table \ref{im}.
The robust parameters were chosen to make the resultant sizes of the synthesized beams of these images to be close to a common resolution.
Then all the images were convolved to a beam with a size of 0\farcs45 for the joint analysis. 
The maximum recoverable angular scales\footnote{From the ALMA Technical Hand Book.} of these observations in the different frequency bands range from 3\farcs4 to 18\farcs7, 
and are all larger than the diameter of the HD~142527 disk (Appendix \ref{allim}). 
Thus, we expect that our results and joint analysis are not affected by the effects of missing fluxes.

\begin{deluxetable}{ccccc}
\tablecaption{Summary of images}
\centering
\tablehead{Continuum/ & Robust & Synthesized & Noise & Channel width \\
Line & & beam & (mJy Beam$^{-1}$) & (km s$^{-1}$)}
\startdata
3 mm & 0 & 0\farcs43 $\times$ 0\farcs36 (77\arcdeg) & 0.013 & \nodata \\
2.1 mm & 2 & 0\farcs41 $\times$ 0\farcs34 (112\arcdeg) & 0.08 & \nodata\\
1.3 mm & 2 & 0\farcs37 $\times$ 0\farcs33 (75\arcdeg) & 0.13 & \nodata \\
0.9 mm & 2 & 0\farcs27 $\times$ 0\farcs21 (81\arcdeg) & 0.3 & \nodata \\
0.6 mm & 2 & 0\farcs39 $\times$ 0\farcs38 (81\arcdeg) & 2.4 & \nodata \\
0.4 mm  & 2 & 0\farcs32 $\times$ 0\farcs25 (121\arcdeg) & 1.1 & \nodata \\
$^{13}$CO (1--0) & $-$0.5 & 0\farcs42 $\times$ 0\farcs35 (74\arcdeg) & 4.3 & 0.05 \\
$^{13}$CO (2--1) & 2 & 0\farcs38 $\times$ 0\farcs33 (75\arcdeg) & 3.2 & 0.1 \\
$^{13}$CO (3--2) & 2 & 0\farcs37 $\times$ 0\farcs32 (112\arcdeg) & 5.4 & 0.1 \\
C$^{18}$O (1--0) & $-$0.5 & 0\farcs43 $\times$ 0\farcs36 (74\arcdeg) & 4.2 & 0.05 \\
C$^{18}$O (2--1) & 2 & 0\farcs38 $\times$ 0\farcs33 (74\arcdeg) & 1.8 & 0.09 \\
C$^{18}$O (3--2) & 2 & 0\farcs27 $\times$ 0\farcs2 (83\arcdeg) & 3 & 0.2 
\enddata
\tablecomments{This table summarizes the original synthesized beams and noise levels of the images. All the images were convolved to a beam of 0\farcs45 for the joint analysis.}
\end{deluxetable}\label{im}

Most of these continuum and molecular-line data have been presented and discussed in detail in the literature \citep{Christiaens14, Casassus15, Muto15, Perez15, Boehler17, Soon19, Francis20, Yamaguchi20}.
Therefore, here we only present a gallery of the images included in our analysis in Appendix \ref{allim}.
A central compact component around the stellar position is detected in the continuum at all the wavelengths. 
We measured the peak position of this component in the continuum images. 
Then we aligned the images obtained in different frequency bands by centering these imaging at the continuum peak positions measured in the corresponding frequency bands.

\section{Analysis and results}
\subsection{Dust distribution}\label{sec_dust}
We performed simple fitting of spectral energy distributions (SED) using the continuum data at six different wavelengths to derive the dust distribution in the protoplanetary disk around HD~142527. 
We assumed the intensity ($I_\nu$) at a given frequency $\nu$ is 
\begin{equation}\label{sed_eq}
I_\nu = (B_\nu(T_{\rm d})-B_\nu(T_{\rm bg}))\cdot(1-e^{-\tau_\nu}), 
\end{equation}
where $B_\nu(T)$ is the Planck function at a temperature of $T$, $T_{\rm d}$ is the dust temperature, $T_{\rm bg}$ is the comic microwave background temperature of 2.73 K, and $\tau_\nu$ is the optical depth at the given frequency, 
and we assumed $\tau_\nu$ to be a simple power-law function of frequency,  
\begin{equation}
\tau_\nu = \tau_{\rm 100~GHz} \times (\frac{\nu}{\rm 100~GHz})^\beta, 
\end{equation}
where $\beta$ is the spectral index of the dust absorption coefficient ($\kappa_\nu$).

\begin{figure*}
\centering
\includegraphics[width=0.9\textwidth]{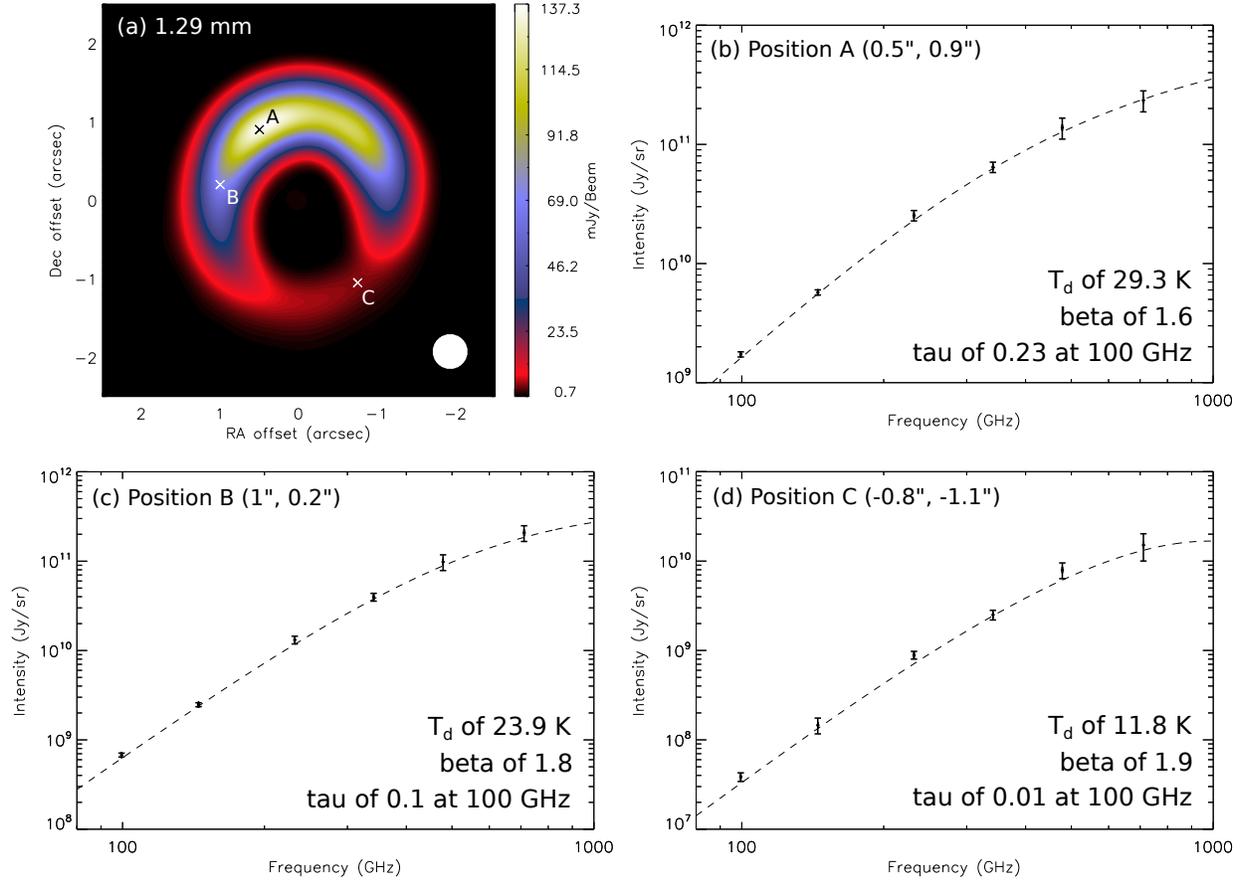}
\caption{Examples of our SED fittings. (a) 1.29 mm continuum image of the HD~142527 disk obtained with the ALMA observations. Three crosses from the north to the south denote the positions of the maximum, median, and minimum intensities of the detected continuum emission at 1.29 mm, respectively. A white filled circle shows the beam size of 0\farcs45. (b)--(d) SEDs at position A, B, and C extracted from the ALMA data at six different wavelengths (data points with error bars). The error bars show the uncertainties due to the noise in the data. Dashed curves present our best-fit SEDs to the observations. The best-fit parameters are listed in the lower right corners in the panels. The offsets with respect to the stellar position of position A, B, and C are labeled in the upper left corners in the panels and also marked in panel (a).}\label{sed}
\end{figure*} 

\begin{figure*}
\centering
\includegraphics[width=0.9\textwidth]{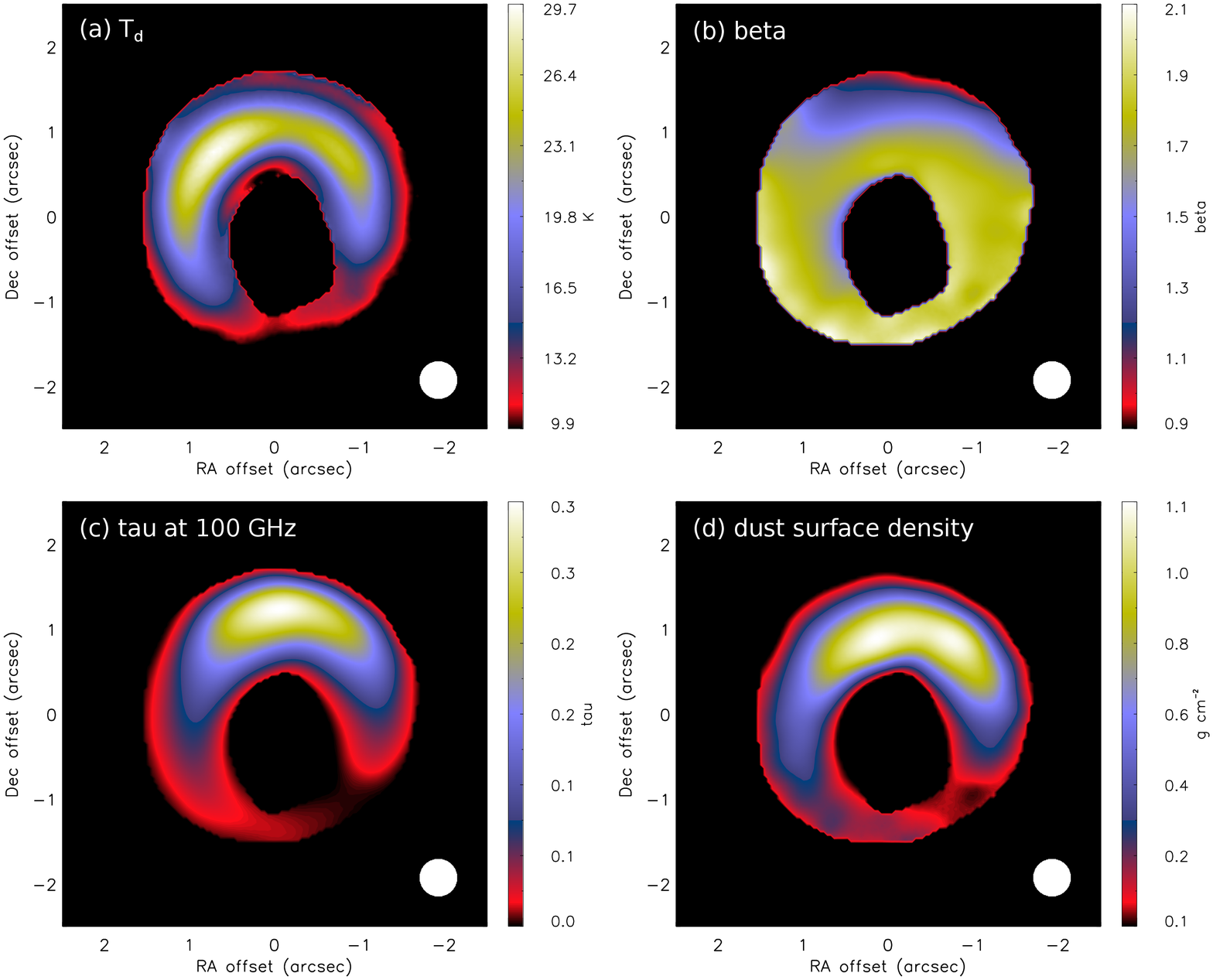}
\caption{Maps of (a) dust temperature ($T_{\rm d}$), (b) spectral index of the dust absorption coefficient ($\beta$), (c) optical depth of the continuum at 3~mm, and (d) dust surface density (in units of g\,cm$^{-2}$) in the HD~142527 disk estimated from our analysis of the SED fitting with the ALMA continuum data at six different wavelengths from 3~mm to 0.4~mm. White filled circles show the beam size of 0\farcs45.}\label{dust}
\end{figure*}

Figure \ref{sed} shows examples of our SED fitting. 
Figure \ref{sed}a presents the continuum image at 1.3 mm. 
A dust ring is clearly seen, and it is brighter in the north.
Figure \ref{sed}b, c, and d show the SEDs at three representative positions in the disk, corresponding to the maximum, median, and minimum intensities of the detected 1.3 mm continuum emission, which are marked with crosses in Fig.~\ref{sed}a.
The best-fit SEDs are shown as dashed curves in these panels and can well describe the observed SEDs within the uncertainties.
The derived maps of $T_{\rm d}$, $\tau_{\rm 100~GHz}$, and $\beta$ are shown in Fig.~\ref{dust}.
The uncertainties range from 10\% to 30\% in $T_{\rm d}$, 15\% to 40\% in $\tau_{\rm 100~GHz}$, and 5\% to 15\% in $\beta$. 
The median uncertainties are 18\%, 22\%, and 8\% in $T_{\rm d}$, $\tau_{\rm 100~GHz}$, and $\beta$, respectively.
The absolute flux uncertainties\footnote{From the ALMA proposer's guide.} of 5\% in Band 3 and 4, 10\% in Band 6 and 7, and 20\% in Band 8 and 9 were considered in our error estimate.

Similar analysis has also been performed in the literature with fewer frequency bands \citep[e.g.,][]{Casassus15, Soon19}.
Our results show a similar trend to that in the literature. 
The spectral indices of $I_{\nu}$ are lower in the northern region, 
and the continuum at wavelengths shorter than 1 mm is optically thick. 
Our derived $\beta$ of $\sim$1.5--1.8 in the northern part of the dust ring is comparable to that derived by \citet{Casassus15}, who included the 8.8 mm data obtained with Australia Telescope Compact Array, 
and it is higher than that derived by \citet{Soon19}, who use the ALMA data at 3 mm and 0.9 mm only.  
This is because the continuum is optically thick at the shorter wavelengths, 
and our study includes the data of the optically thin continuum at three longer wavelengths from 1.3 to 3 mm. 
Our derived $\beta$ is primarily determined by the data at these three wavelengths and is least affected by the opacity effect.

We then estimated the dust surface density from the derived $\beta$ and $\tau_{\rm 100~GHz}$. 
We adopted the function of the dust absorption coefficient in \citet{Beckwith90}, 
\begin{equation}
\kappa_\nu=10\times(\frac{\nu}{\rm 1000~GHz})^\beta~{\rm g\,cm^{-2}}.
\end{equation}
The derived dust density map is shown in Fig.~\ref{dust}d. 
The uncertainties in the estimated surface density range from 30\% to 90\% with a median uncertainty of 42\%.
Our results suggest that the dust is more concentrated toward the northwestern part of the dust ring, 
which is different from the intensity distribution.
This general trend is consistent with the previous results in the literature \citep{Soon19}, 
but the estimated values and the contrast between the dust densities in the northern and southern parts of the dust ring are different, 
which is because our analysis on the six frequency bands has better constraints on the dust temperature and opacity. 
We note that the values of the estimated dust surface density in Fig.~\ref{dust} are subject to the adopted dust absorption coefficient \citep{Birnstiel18}. 

We also note that dust scattering becomes important at shorter wavelengths in the northern part of the dust ring, as revealed by the ALMA observations of the polarized continuum emission at millimeter wavelengths \citep{Kataoka16, Ohashi18}. 
The effects of dust scattering were not considered in our simple SED fitting.
The contribution of the scattering opacity to observed intensity depends on the optical depth of the continuum \citep{Liu19, Zhu19}. 
If the continuum is optically thin, the observed intensity is approximately the same as the expected intensity calculated without considering dust scattering for a given temperature. 
In contrast, when the continuum is optically thick, the observed intensity becomes lower due to dust scattering. 
In addition, if the continuum in a disk is optically thick along the radial direction and optically thin along the vertical direction, 
the observed intensity can become higher than the expected intensity calculated without considering dust scattering \citep{Ueda20}. 
In the study of the SED at the center of the disk around TW~Hya, 
\citet{Ueda20} showed that after incorporating dust scattering in their model calculations, compared to the case without dust scattering, the computed intensity at $\sim$1~mm becomes 35\% lower and that at $>$3~mm becomes 20\% higher, when the optical depth at 3~mm is 0.95 and the dust temperature is 30~K at 10 au in the model disk. 
Comparing our SED calculations to those in \citet{Ueda20}, 
the dust temperature at the location of the intensity peak in the HD~142527 disk is comparable, 
but the optical depth at 3~mm in the HD~142527 disk is a factor of four lower. 
The continuum in the HD~142527 disk becomes optically thick only at wavelengths shorter than 1~mm. 
In addition, similar to the TW~Hya disk, the HD~142527 disk is also close to face-on with an inclination angle of 26$\arcdeg$ (Section~\ref{kinematics}). 
Therefore, 
because of the low optical depth of the continuum in the HD~142527 disk, 
we expect that in the HD~142527 disk, the reduction of the observed intensity due to dust scattering is less than 35\% at wavelengths shorter than 1~mm, 
and there is no effect due to dust scattering at the longer wavelengths. 
We have tested our SED fitting at the 1.3 mm intensity peak by artificially increasing the intensities at 0.9 mm to 0.4 mm by 20\% to 30\%. 
The best-fit $T_{\rm d}$ and $\tau_{\rm 100~GHz}$ become higher and lower by 10\% to 30\%, respectively, and $\beta$ decreases by 0.1. 
This change in $\beta$ is less than 10\%. 
These values are comparable to the uncertainties of our fitting results. 
Thus, the effects of dust scattering (if any) on our SED analysis are not significant, 
and our derived $T_{\rm d}$, $\tau_{\rm 100~GHz}$, and $\beta$ maps are still valid even though the effects of dust scattering are not included in our SED calculations.

\subsection{Gas distribution}\label{gas}
\begin{figure}
\centering
\includegraphics[width=0.48\textwidth]{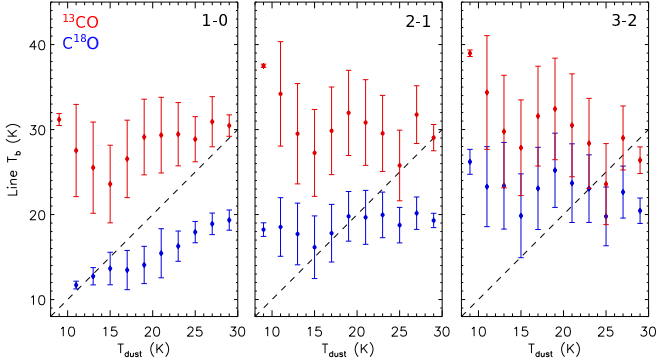}
\caption{Comparison of the peak brightness temperature ($T_{\rm b}$; after continuum subtraction) of the $^{13}$CO (red data points) and C$^{18}$O (blue data points) lines of $J = 1\mbox{--}0$ (left panel), 2--1 (middle panel), and 3--2 (right panel) with the estimated dust temperature in the HD~142527 disk. Vertical and horizontal axes show the peak brightness temperature of the lines and the dust temperature, respectively. The data were extracted for every pixel in the disk and then were binned up with a bin size of $T_{\rm d}$ of 2~K. The data points and the error bars in these panels present the means and 1$\sigma$ dispersions of $T_{\rm b}$ for every $T_{\rm d}$ bins.}\label{tb}
\end{figure}

To select the suitable molecular lines for estimation of the gas distribution in the protoplanetary disk around HD~142527, 
we first compared the peak brightness temperature (after continuum subtraction; $T_{\rm b}$) with the estimated dust temperature for every position in the disk (Fig.~\ref{tb}).
For clarity of presentation, 
we binned up the data points with a bin size of $T_{\rm d}$ of 2~K.
Figure~\ref{tb} shows the means and 1$\sigma$ dispersions of $T_{\rm b}$ of different lines as a function of $T_{\rm d}$.
On assumptions of typical density and temperature profiles along the vertical direction in a protoplanetary disk \citep[e.g.,][]{Pinte18}, 
where the density is higher in the midplane and the temperature is higher at the surface, 
the peak brightness temperature of a molecular line is expected to be lower than the midplane temperature when the line is optically thin. 
The brightness temperature is higher than the midplane temperature if the line is already optically thick in an upper layer of the disk. 
Assuming the dust grains have settled to the midplane and are thermally coupled with gas in the HD~142527 disk, 
our estimated $T_{\rm d}$ in Fig.~\ref{dust} can represent the midplane temperature. 
In Fig.~\ref{tb}, $T_{\rm b}$ of all the lines except C$^{18}$O (1--0) is higher than $T_{\rm d}$. 
Although $T_{\rm b}$ of some of those lines becomes lower than $T_{\rm d}$ when $T_{\rm d}$ is high,  
this is most likely due to the continuum over subtraction, as discussed in the previous studies \citep[e.g.,][]{Casassus15, Boehler17, Soon19}.
Therefore, the comparisons between $T_{\rm b}$ and $T_{\rm d}$ suggest that only the C$^{18}$O (1--0) line is likely optically thin.
In addition, $T_{\rm b}$ of the $^{13}$CO (3--2) and (2--1) lines is higher than that of $^{13}$CO (1--0) as well as C$^{18}$O (3--2) and (2--1). 
These lines are optically thick. 
Thus, the $^{13}$CO lines likely become optically thick in upper layers, while the C$^{18}$O (3--2) and (2--1) lines becomes optically thick in layers closer to the midplane, and there is a vertical temperature gradient, similar to other protoplanetary disks \citep[e.g.,][]{Pinte18}. 

Therefore, we adopted the data of the three C$^{18}$O lines to estimate the gas distribution in the HD~142527 disk because the C$^{18}$O (1--0) line is possibly optically thin.
To consider the vertical structures in the disk in our estimation, 
we adopted the vertical temperature and density profiles derived in \citet{Lee15}. 
At a given radius, the vertical temperature profile ($T_z$) was assumed to be 
\begin{equation}
T_z = T_{\rm m} \times \left \{
\begin{array}{lc}
\{1 + (1/2)A[1-\cos\pi (z/Z_{\rm a})^n]\} & \mbox{if } z < Z_{\rm a}\\
(1 + A) &\mbox{if } z \geq Z_{\rm a}, \\ 
\end{array}
\right.,
\end{equation}
where $z$ is the vertical distance from the midplane, $A$ is defined as $(T_{\rm a} - T_{\rm m})/T_{\rm m}$, $T_{\rm m}$ is the temperature in the disk midplane, $T_{\rm a}$ is the temperature in the atmosphere of the disk, and $Z_{\rm a}$ is the transitional height to the isothermal atmosphere. 
This profile is similar to those adopted in the other observational studies \citep[e.g.,][]{Dartois03,Rosenfeld13}.
Here we adopted $n=2$.
As derived in \citet{Lee15}, with $n=2$, the hydrostatic equilibrium can be solved analytically.
The vertical pressure profile ($P_z$) in the hydrostatic equilibrium is 
\begin{equation}
P_z = P_{\rm m} \times \left \{
\begin{array}{lc}
e^{\{-\frac{{Z_{\rm a}}^2}{\pi \sqrt{1+A}}\arctan[\sqrt{1+A}\tan\frac{\pi}{2}(\frac{z}{Z_{\rm a}})^2] \}} & \mbox{if } z < Z_{\rm a}\\
e^{[-\frac{{Z_{\rm a}}^2}{\pi \sqrt{1+A}}-\frac{z^2-{Z_{\rm a}}^2}{2(1+A)}]} &\mbox{if } z \geq Z_{\rm a}, \\ 
\end{array}
\right. ,
\end{equation}
where $P_{\rm m}$ is the pressure in the disk midplane. 
Then the vertical density profile is computed as $P_z$/$T_z$ \citep{Lee15}.
In our calculations, 
$T_m$ was adopted to the dust temperature. 
$T_{\rm a}$ was adopted to be the peak brightness temperature of $^{13}$CO (3--2) without continuum subtraction, because the $^{13}$CO (3--2) line is most optically thick among the lines analyzed in this work and likely traces the disk surface. 
$Z_{\rm a}$ was assumed to be the pressure scale height\footnote{We have also tried $Z_{\rm a}$ of 1.5 or 2 times the pressure scale height and found that adopting $Z_{\rm a}$ to be one pressure scale height provides better fits to the observed spectra.}, 
computed with the midplane temperature and the stellar mass of 2.48~$M_\odot$ (Section~\ref{kinematics}).
 
\begin{figure*}
\centering
\includegraphics[width=0.9\textwidth]{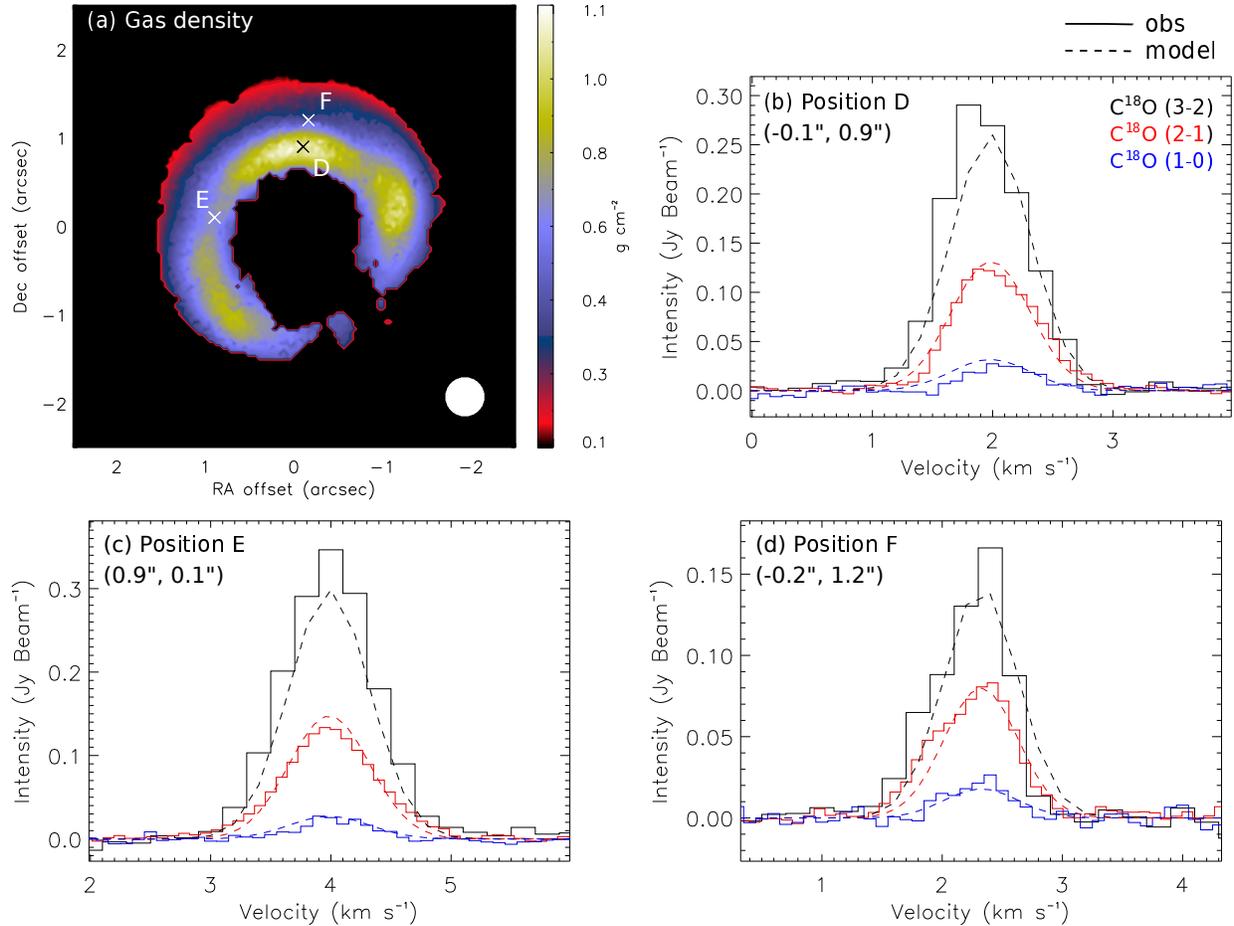}
\caption{(a) Gas surface density (in units of g~cm$^{-2}$) in the HD~142527 disk estimated from the model fitting to the C$^{18}$O (3--2), (2--1), and (1--0) spectra. A white filled circle shows the beam size of 0\farcs45. (b)--(d) Examples of our model fitting to the C$^{18}$O (3--2; black histograms), (2--1; red histograms), and (1--0; blue histograms) spectra at position D, E, and F. Solid histograms and dashed curves show the observed and model spectra, respectively. The offsets with respect to the stellar position of position D, E, and F are labeled in the upper left corners in the panels and are also denoted in panel (a). Position D, E, and F correspond to the positions of the maximum and median gas surface densities and the lowest gas-to-dust mass ratio in the HD~142527 disk, respectively.}\label{gasmap}
\end{figure*}  
 
With the known $T_{\rm m}$, $T_{\rm a}$, and $Z_{\rm a}$ and the assumed vertical temperature and density profiles, 
we computed the radiative transfer equation assuming the local thermal equilibrium (LTE) condition, 
integrated the equation along the line of sight, 
and generated model spectra of the C$^{18}$O (3--2), (2--1), and (1--0) lines to fit the observations for every position in the HD~142527 disk.
We did not consider that the line of sight is inclined with respect to the disk midplane in our radiative transfer calculation because the HD~142527 disk is close to face on with an inclination angle of 26$\arcdeg$ (Section \ref{kinematics}), 
For a given position, the centroid velocity and line width of the model spectra were adopted to be the mean centroid velocity and line width of the three C$^{18}$O lines, which were measured by fitting Gaussian profiles to the observed spectra. 
The C$^{18}$O abundance was assumed to be a constant of  $2\times10^{-7}$ \citep{Wilson94, Jorgensen04}, 
and it was set to be zero in the regions where the temperature is below 20 K to mimic the effects of CO frozen out. 
In addition, we included a geometrically thin dust layer in the midplane, which is segregated from the gas components, in our radiative transfer calculation. 
Thus, our calculations also took the continuum subtraction and the suppression of the line intensity by the continuum into account.
The dust temperature and opacity in our dust layer were adopted from the results of the SED fitting in Section~\ref{sec_dust}.

Figure \ref{gasmap} presents the map of the estimated gas surface density in the HD~142527 disk and examples of our model fitting to the C$^{18}$O (3--2), (2--1), and (1--0) spectra. 
The spectra at three representative positions are shown, which correspond to the positions with the maximum and median gas surface densities and the lowest gas-to-dust mass ratio in the HD~142527 disk.
The fluxes of our model spectra typically match with the observations within the 3$\sigma$ uncertainties, or the difference is less than 20\%. 
The uncertainties in the estimated gas surface density range from 8\% to a factor of two, 
and the median uncertainty is 22\%.
The uncertainties in $T_{\rm m}$, $T_{\rm a}$, continuum optical depths, and absolute flux calibration were all included in our error propagation. 
Figure \ref{gasmap}a shows that the gas density is highest in the northern part of the disk, similar to the dust density distribution. 
In addition, there are two local maximums of the gas density, one in the west and one in the southeast. 
These locations correspond to the regions showing lower $^{13}$CO (3--2) peak brightness temperatures (see Fig.~4 in \citet{Soon19}). 
Our radiative transfer calculations also show that the C$^{18}$O (1--0) line is optically thin to marginally optically thick, where the optical depth can reach one to two.

\subsection{Gas-to-dust mass ratio}\label{goverd}
\begin{figure}
\centering
\includegraphics[width=0.48\textwidth]{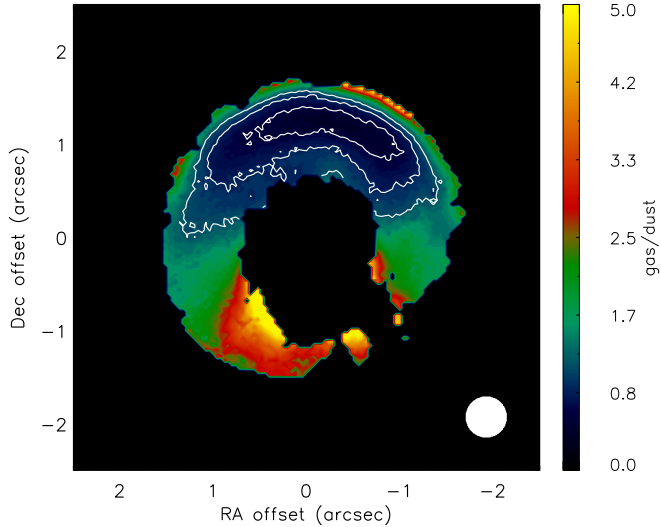}
\caption{Gas-to-dust mass ratio in the HD~142527 disk estimated from the dust (Fig.~\ref{dust}c) and gas (Fig.~\ref{gasmap}a) surface densities. The estimated ratios range from 0.4 to 9. The inner to outer contours delineate the regions with the gas-to-dust ratios of 0.6, 0.9 and 1.2. The C$^{18}$O abundance is assumed to be a constant of $2\times10^{-7}$ in the entire disk. We note that the values of the estimated gas-to-dust mass ratio are subject to the actual C$^{18}$O abundance and dust absorption coefficient as a function of frequency. }\label{gd}
\end{figure}

With the gas and dust surface density estimated from the C$^{18}$O lines and the continuum, 
the gas-to-dust mass ratio was computed. 
The results are shown in Fig.~\ref{gd}.
The estimated gas-to-dust mass ratios range from 0.4 to 9 with a mean value of 1.6, 
and the ratios are lower in the north and gradually increase toward the south in the disk.
The uncertainties in the estimated gas-to-dust mass ratio range from 30\% to a factor of 1.3 with a median uncertainty of 50\%. 
In the northern region, which is delineated by the inner contour in Fig.~\ref{gd}, the gas-to-dust mass ratio is more than a factor of three lower than the mean ratio in the disk. 
Thus, the gas-to-dust mass ratio is indeed significantly lower in the north in the disk.
This trend of the lower gas-to-dust mass ratios in the north than in the south has also been reported by \citet{Soon19}.

We note that the abundances of CO isotopologues are uncertain in protoplanetary disks \citep{Furuya14, Miotello16}, 
and the actual C$^{18}$O abundance can be orders of magnitude lower than the values in the interstellar medium \citep[e.g.,][]{McClure16,Schwarz16}. 
If the C$^{18}$O abundance is ten times lower than our assumed value, 
the derived gas-to-dust mass ratio becomes ten times higher.
In addition, the dust absorption coefficient can also be uncertain \citep[][]{Birnstiel18}. 
Therefore, in the present paper, we do not discuss the absolute values of the estimated gas-to-dust mass ratio, 
and our discussions focus on the trend in the distribution of the gas-to-dust mass ratio in the HD~142527 disk, 
which is valid if the dust absorption coefficient and the C$^{18}$O abundance are more or less uniform in the disk.

\subsection{Gas kinematics}\label{kinematics}
We measured the disk rotation and orientation by fitting the molecular-line data with Keplerian disk models using {\it DiskFit} \citep{Pietu07}, which is a package built in the software GILDAS\footnote{\url{http://www.iram.fr/IRAMFR/GILDAS}} \citep{Pety05, gildas13}.
With {\it DiskFit}, we constructed Keplerian disk models which have power-law density and temperature profiles and conventional vertical structures. 
The detail descriptions of the disk models are in \citet{Pietu07}.
Then radiative transfer calculations of the model disks were performed, the model images were Fourier transformed and sampled with the same {\it uv} coverage as the observations, and the fitting to the observational data was done in the {\it uv} domain, in {\it DiskFit}. 

\begin{deluxetable*}{cccccccccccccc}
\tablecaption{Best-fit parameters with the C$^{18}$O (1--0) and (2--1) data}
\centering
\tablehead{Line & $M_\star$ & $i$ & $PA$ & $V_{\rm sys}$ & $N_0$ & $p$ & $T_0$ & $q$ & $R_{\rm in}$ & $R_{\rm out}$ & $H_0$ & $h$ & $\Delta V$\\
& ($M_\odot$) & & & (km s$^{-1}$) & (cm$^{-2}$) & & (K) & & (au) & (au) & (au) & & (km s$^{-1}$)}
\startdata
C$^{18}$O (1--0) & 2.48 & 26.1\arcdeg & 340.3\arcdeg & 3.74 & 10$^{17.1}$ & $-$2.2 & 19 & $-$0.4 & 81 & 286 & 7.4 & 1.3 & 0.19 \\
C$^{18}$O (2--1) & 2.48 & 26.7\arcdeg & 340.8\arcdeg & 3.74 & 10$^{17.0}$ & $-$3.4 & 33 & $-$0.4 & 121 & 387 & 9 & 1.3 & 0.29
\enddata 
\tablecomments{$M_\star$, $i$, $PA$, and $V_{\rm sys}$ are the stellar mass, inclination angle, position angle of the major axis, and systemic velocity of the model disks, respectively. $R_{\rm in}$ and $R_{\rm out}$ are the inner- and outermost radii of the model disks. In the disk models, radial profiles of C$^{18}$O column density, gas temperature, and pressure scale height are assumed to be power-law functions with power-law indices of $p$, $q$, and $h$, respectively, and $N_0$, $T_0$, and $H_0$ present the values at a radius of 100 au in the midplane. $\Delta V$ is the line width in the disk models.  The uncertainties in the best-fit parameters due to the the noise in the data are less than 5\%. The detailed descriptions of the disk models are in \citet{Pietu07}.}
\end{deluxetable*}\label{bestfit}

\begin{figure*}
\centering
\includegraphics[width=0.95\textwidth]{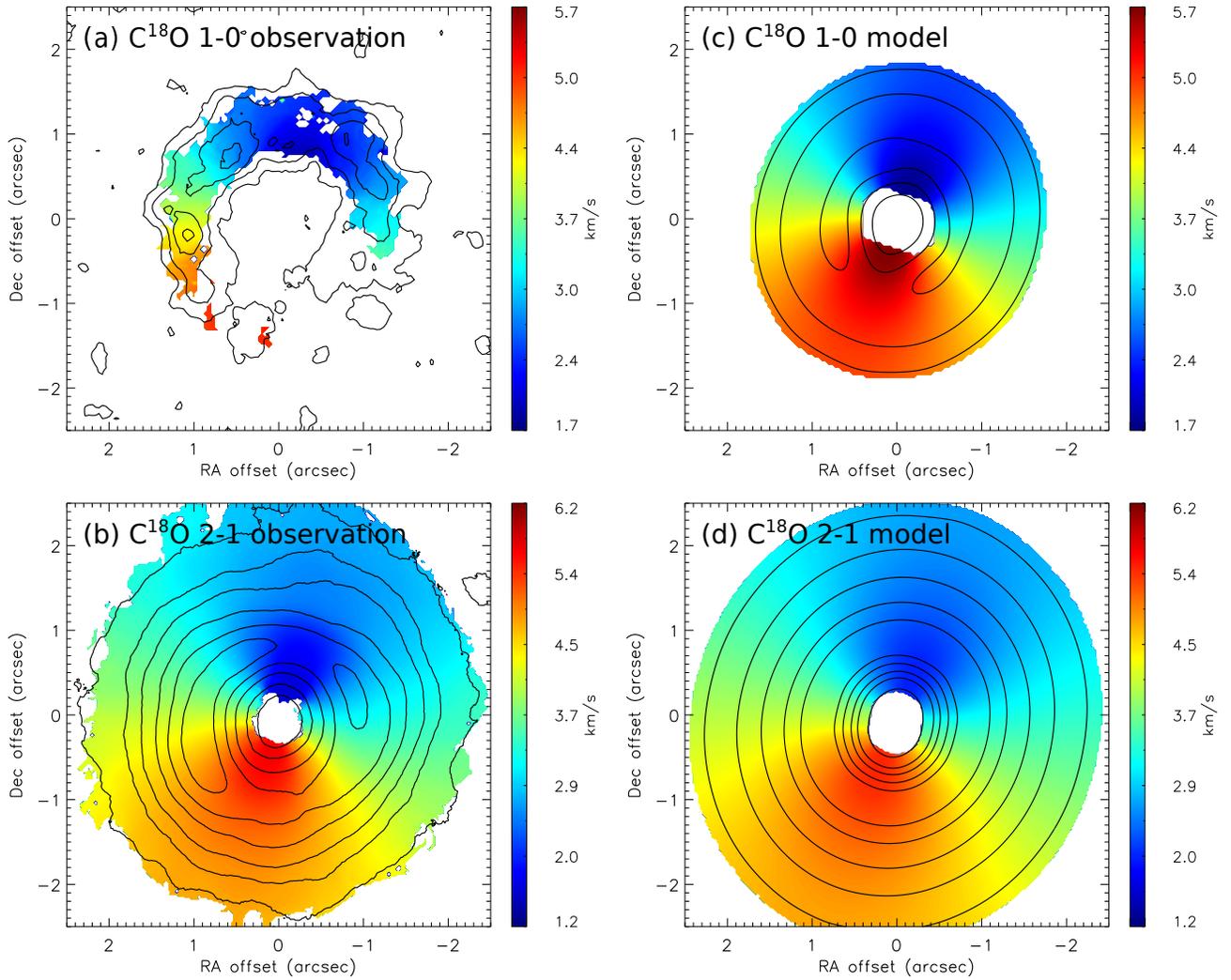}
\caption{Comparison of the total integrated intensity (moment 0; contours) and peak velocity (in units of km\,s$^{-1}$; color scale) maps of the C$^{18}$O (1--0; panel a) and (2--1; panel b) emission observed in the HD~142527 disk with those of our best-fit model disks (panel c and d). The model images were generated with radiative transfer calculations and sampled with the same {\it uv} coverages as the observations. All the maps were convolved to the same beam size of 0\farcs45. Contour levels in (a) and (c) start from 3$\sigma$ in steps of 3$\sigma$, where 1$\sigma$ is 1.8 mJy~Beam$^{-1}$~km~s$^{-1}$, and those in (b) and (d) are 5$\sigma$, 15$\sigma$, 25$\sigma$, 40$\sigma$, 55$\sigma$, and 70$\sigma$, where 1$\sigma$ is 1.7 mJy~Beam$^{-1}$~km~s$^{-1}$.}\label{model}
\end{figure*} 

In this work, we performed the fitting on the C$^{18}$O (1--0) and (2--1) data. 
These two lines have lower optical depths compared to the other CO isotologue lines and trace layers closer to the midplane in the HD~142527 disk (Section \ref{gas}).
Thus, the fitting of these two lines has least uncertainties due to not well constrained heights of the emitting layers. 
The fitting of the two lines was performed separately for a consistency check. 
The best-fit parameters are listed in Table~\ref{bestfit}.
Figure~\ref{model} presents the synthetic maps of the intensity and peak velocity of our best-fit Keplerian disk models in comparison with the observed maps. 

We note that our model disks are axisymmetric and the observed disk is asymmetric, 
so that there are significant residuals after subtracting the model intensity distributions from the observations. 
Nevertheless, the fitting results of the C$^{18}$O (1--0) and (2--1) lines are consistent, even though the two lines exhibit very distinct intensity distributions (Fig.~\ref{model}a and b). 
The estimated stellar mass and systemic velocity from the fitting of the two different lines are identical, 
and the difference in the estimated position and inclination angles is less than 1$\arcdeg$.
In addition, our measured stellar mass and inclination and position angles of the disk are also consistent with the previous results using the ALMA $^{13}$CO and C$^{18}$O (3--2) data at a resolution of $\sim$0\farcs2--0\farcs4 \citep{Fukagawa13,Boehler17}.
Therefore, the disk rotation is properly measured, and the measurements are not affected by the asymmetric intensity distributions in the observations, 
although the observed intensity distributions cannot be fully reproduced with the models. 

We also note that HD~142527 is a binary system with a mass ratio of 0.1--0.2 and a separation of 0\farcs1 on the plane of the sky \citep{Christiaens18}.
The C$^{18}$O (1--0) emission shows a ring-like structure at a radius of 1\arcsec. 
The C$^{18}$O (2--1) emission extends out to more than 2$\arcsec$ and exhibits an inner cavity with a radius of approximately 1$\arcsec$.
Thus, the majority of the emission of these two lines is distributed ten times away from the binary system.  
We expect that the disk rotation traced by the C$^{18}$O (1--0) and (2--1) lines can be approximated with the Keplerian rotation around a central point mass, and our estimated $M_\star$ represents the total mass of the binary system.

\section{Discussion}
\subsection{Deviation from Keplerian rotation}
\begin{figure*}
\centering
\includegraphics[width=0.45\textwidth]{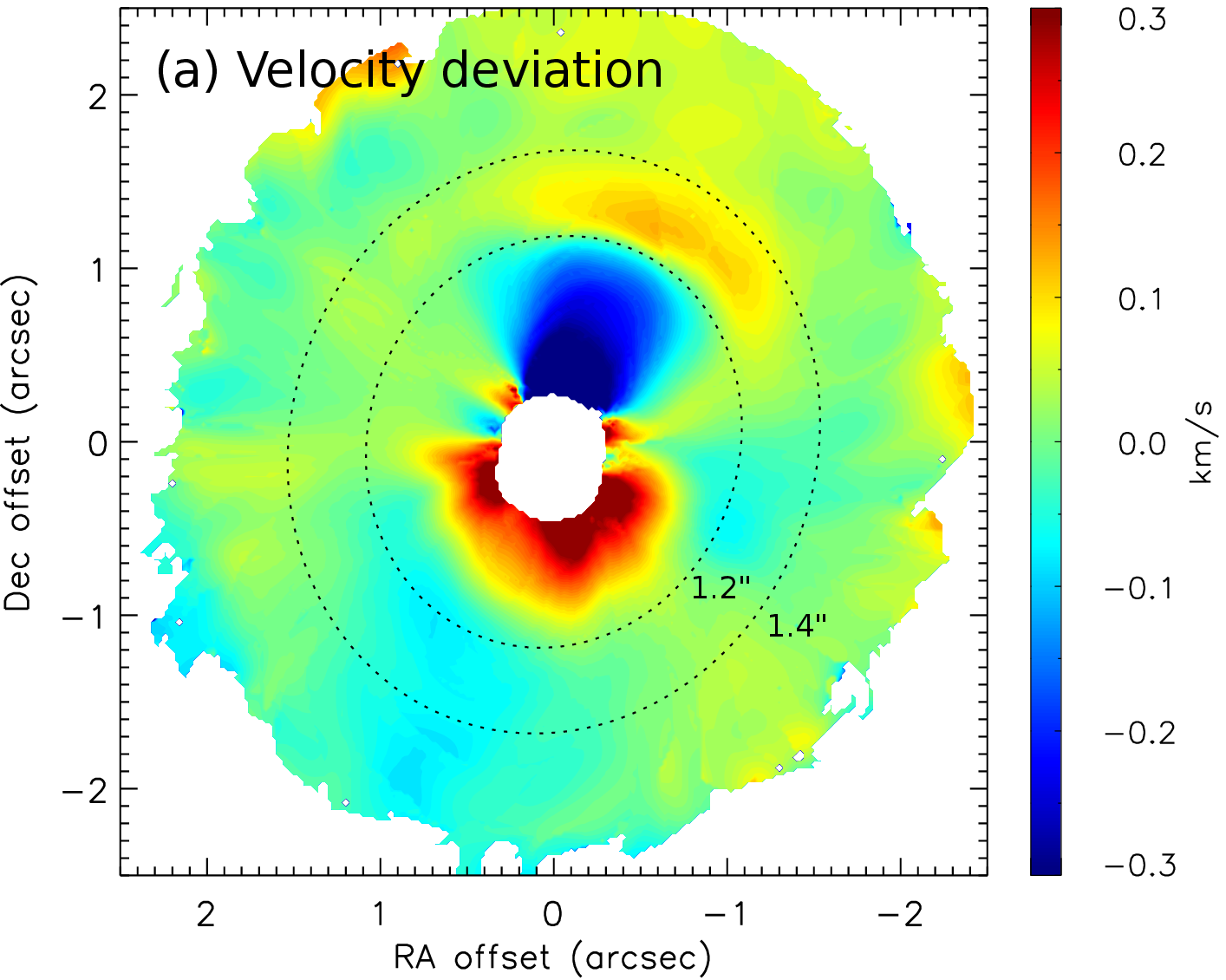}
\includegraphics[width=0.45\textwidth]{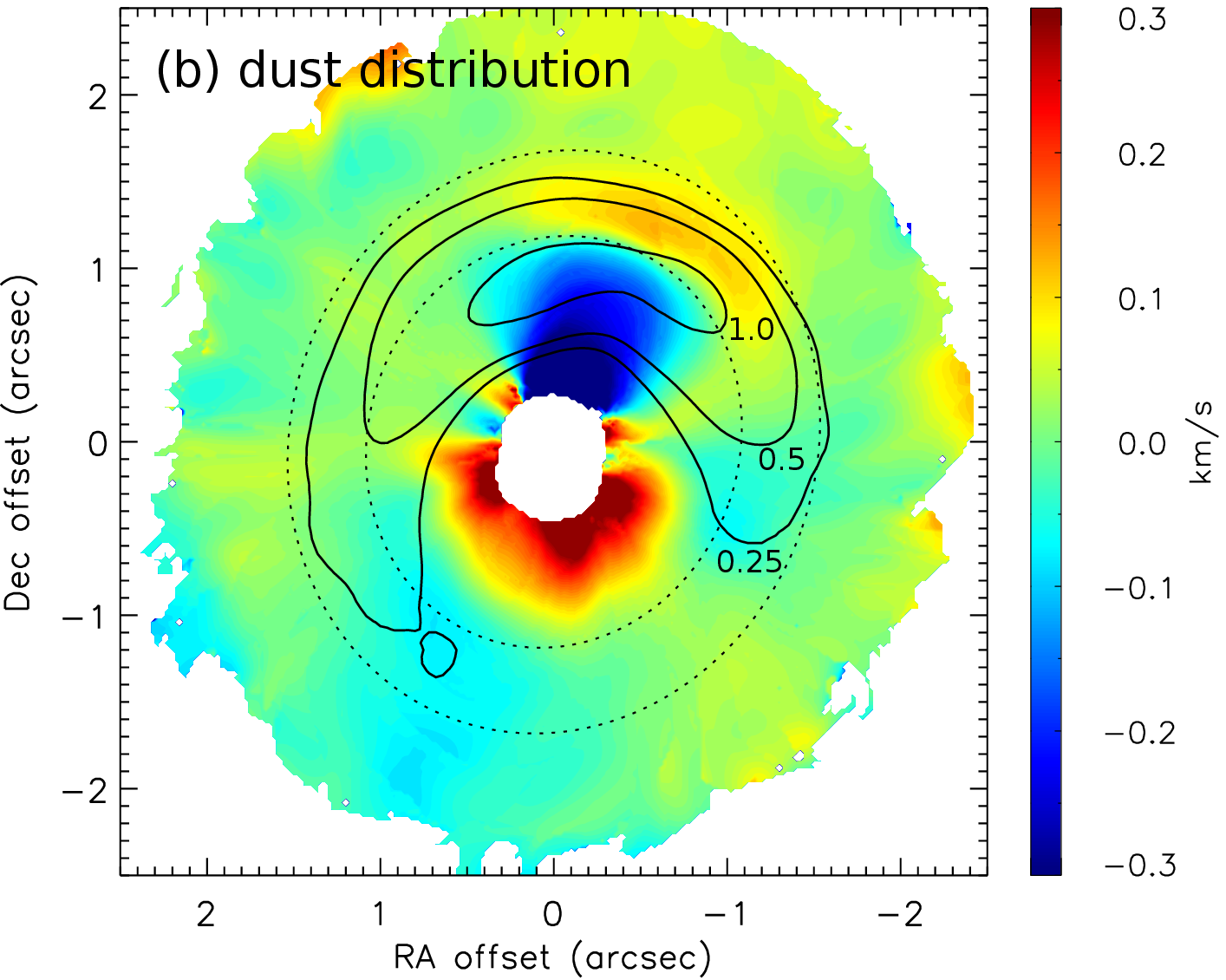}
\includegraphics[width=0.45\textwidth]{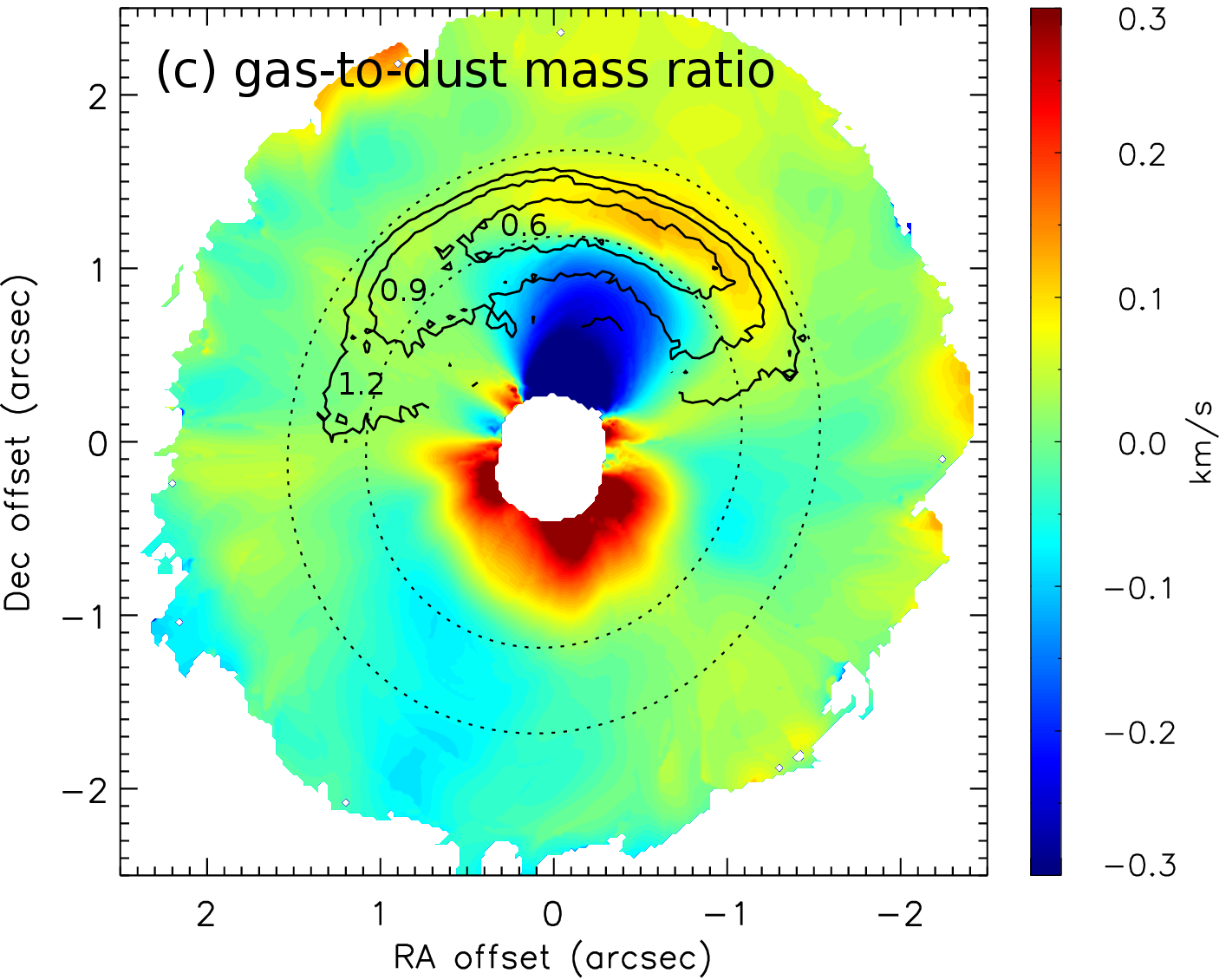}
\includegraphics[width=0.45\textwidth]{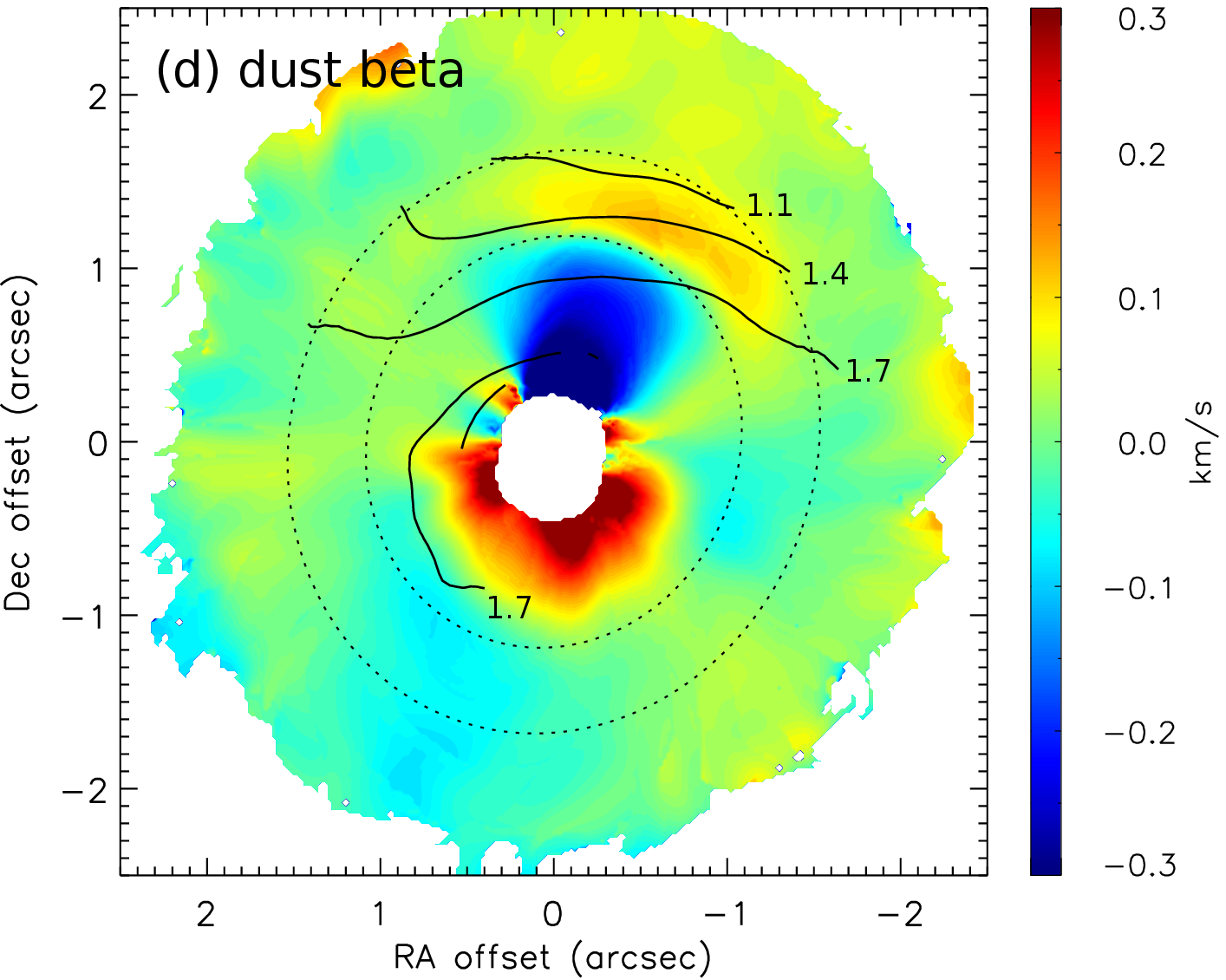}
\caption{Color scale maps show the velocity deviation from the circular Keplerian rotation detected in the C$^{18}$O (2--1) emission in the HD~142527 disk (in units of km\,s$^{-1}$). The velocity resolution of the C$^{18}$O (2--1) data is 0.09 km~s$^{-1}$. The velocity deviation in the central 0\farcs5 region could be related to the warped accretion flow around the binary system \citep{Christiaens18}, which is not the scope of the present paper due to our limited angular resolution of 0\farcs45. In our discussion, we focus on the velocity deviation in the northwest at position angle from 300$\arcdeg$ to 360$\arcdeg$ and radii close to 1\farcs2. Panel b--d present the comparison of the velocity deviation (color scale) with the dust surface density (contours in panel b), the gas-to-dust mass ratio (contours in panel c), and the spectral index of dust absorption coefficient ($\beta$; contours in panel d)  in the HD~142527 disk. Dotted inner and outer ellipses in each panel denote radii of 1\farcs2 and 1\farcs4 in the disk midplane.}\label{dv}
\end{figure*}

To investigate possible perturbation in the protoplanetary disk around HD~142527 due to the interaction between the disk and the binary system, 
we subtracted the peak velocity map of our best-fit Keplerian disk model of the C$^{18}$O (2--1) line (Fig.~\ref{model}d) from the observed velocity map (Fig.~\ref{model}b).
The residual velocity map is shown in Fig.~\ref{dv}a. 
In the residual velocity map, there are clear blue- and redshifted excesses at radii smaller than 1\farcs2 in the northwest and southeast, respectively, 
which are along the major axis of the disk.
At radii between 1\farcs2 and 1\farcs4, there is redshifted excess with a maximum residual velocity of 0.13~km\,s$^{-1}$ in the northwest. 
The signal-to-noise ratios of the C$^{18}$O (2--1) emission are higher than 40 at these positions, 
and thus the uncertainties in the measured peak velocity due to the noise are negligible compared to the velocity channel width of 0.09 km s$^{-1}$.
In the southeastern side of the disk, 
there is possible blueshifted excess, 
but the residual velocity is less than one channel width.
Other regions along the minor axis also do not exhibit significant residual velocity more than the channel width, except for those very close to the edges of the map, where the peak velocity is uncertain due to fainter emission.

To examine whether these residual velocities are caused by the errors in our disk model, 
we generated a series of the residual velocity maps due to different types of the errors in the model (Appendix \ref{simdv}).
The patterns in the observed residual velocity map are different from those in the simulated residual velocity maps assuming that there are errors in the disk center, position angle of the disk major axis, inclination angle, or height of the emission emitting layer in the disk models.
Therefore, the observed residual velocity map most likely traces the deviation of the gas motion from the circular Keplerian rotation in the HD~142527 disk projected onto the line of sight, and is not due to the accuracy of our disk model. 
We also compared the observed velocity deviation with the expected velocity deviation due to non-zero eccentricity in a disk \citep{Goodchild06,Hsieh12} because the dust ring in the HD~142527 is possibly eccentric \citep{Boehler17}.
Nevertheless, the observed velocity deviation is inconsistent with the expectation for an eccentric disk (Appendix \ref{simdv}). 

We note that the angular resolution of 0\farcs45 of our molecular-line maps is limited, and any velocity feature within a radius of 0\farcs45 is uncertain due to the beam convolution. 
In addition, the velocity field in the central region within a radius of 0\farcs5 is complicated as revealed in the CO (6--5) emission at a high angular resolution of $<$0\farcs1, which could be due to warped accretion flows \citep{Casassus15a}. 
Thus, in the present work, our discussion focuses on the velocity deviation at radii larger than 0\farcs5. 

\subsection{Kinematical signs of a pressure bump}
In Fig.~\ref{dv}, 
in addition to the velocity deviation around the disk center, 
the most significant velocity deviations appear at position angles from 300$\arcdeg$ to 360$\arcdeg$ and inside and outside a radius of 1\farcs2, which is approximately the radius of the dust ring. 
The position angle of the major axis of the HD~142527 disk is 340$\arcdeg$. 
These velocity deviations are at position angles close to the major axis, 
where any radial motion has minimal contribution to the line-of-sight velocity.   
Considering projection effects, the observed velocity deviations are more likely caused by azimuthal gas motions.  
In the HD~142527 disk, the disk rotation induces the blueshifted emission in the northwestern side of the disk. 
Thus, the blue- and redshifted velocity excesses in the northwest suggest that the gas motions are faster and slower than the Keplerian rotation, respectively. 
This feature, super- and sub-Keplerian rotation inside and outside the dust ring, is consistent with the expectation from the existence of a pressure bump at a radius of 1\farcs2 in the northwest \citep{Perez18, Teague18, Rosotti20}. 
When a pressure bump is present, inside and outside a pressure bump, there are additional forces pulling inward and pushing outward due to the pressure gradient, respectively.
As a result, disk rotation inside and outside a pressure bump becomes super- and sub-Keplerian rotation, respectively. 
For comparison, in Appendix~\ref{simdv}, we also computed a residual velocity map due to an axisymmetric pressure bump at a radius of 1\farcs2. 
The observed velocity pattern is indeed similar to that in the northwest in the model residual velocity map (Fig.~\ref{dv_im}e).
Nevertheless, the observed velocity deviation is asymmetric, 
which is different from the model residual velocity map of an axisymmetric pressure bump in Appendix~\ref{simdv}, 
and there is no corresponding feature in the southeast in the observed map.
Thus, in the HD~142527 disk, the radial profile of the gas pressure is possibly not axisymmetric, 
and there could be a stronger pressure gradient in the northwest than the southeast.

We note that the presence of the local pressure bump in the northwest suggested by the gas kinematics is not seen in the distribution of the gas pressure derived from our analysis on the gas density and temperature in Section \ref{gas}.
In our analysis, the gas density and temperature were estimated from the C$^{18}$O intensity distributions.
The variations in the intensity distributions due to this local pressure bump are possibly smoothed out due to the beam convolution, 
and thus do not appear in our derived pressure distribution. 
On the contrary, the peak velocity patterns are less sensitive to the beam convolution compared to the intensity distributions because the intensity peaks at different positions in the disk are at different velocities.

\subsection{Dust trapping by pressure bump}\label{trapping}
Figure \ref{dv}b compares the dust distribution and the velocity deviation from the Keplerian rotation in the HD~142527 disk. 
The dust distribution is concentrated in the northwest in the disk, where the dust surface density is two to three times larger than the mean value of the disk,  
and this dust concentration is approximately coincident with the region showing the velocity deviation due to the possible pressure bump. 
Figure \ref{dv}c compares the distribution of the gas-to-dust mass ratio with the velocity deviation. 
The gas-to-dust mass ratio gradually decreases toward the region of the pressure bump. 
In the pressure bump, the gas-to-dust mass ratio becomes a factor of three lower than the mean ratio in the disk.
Therefore, our results show a high dust surface density as well as a low gas-to-dust mass ratio in the region exhibiting the kinematical features of a pressure bump in the HD~142527 disk. 
These results suggest dust trapping by a pressure bump in the northwest in the HD~142527 disk.

\subsection{Origins of the pressure bump and dust trapping}
Theoretical studies show that dust grains can be trapped by a pressure maximum of a vortex formed at the edge of a cavity in a protoplanetary \citep[e.g.,][]{Lin14,Zhu14,Baruteau16},
which has been proposed to explain asymmetric, horseshoes-like continuum emission in protoplanetary disks \citep[e.g.,][]{Perez14,Marel15,Marel16,Cazzoletti18,Baruteau19}.
A vortex is expected to have an anticyclonic motion with a velocity on the order of 0.1 km s$^{-1}$ and exhibit a velocity gradient along the azimuthal direction \citep{Huang18,Perez18,Robert20}.
However, due to the limited resolutions, 
we are not able to detect the azimuthal velocity gradient of the anticyclonic motion expected from a vortex projected onto the line of sight, 
even if the pressure bump which we detected is indeed caused by a vortex. 

The other possible origin of the observed pressure bump is the spiral driven by the massive companion of HD~142527.
The hydrodynamical simulations for the HD~142527 disk show that the companion of HD~142527 with a mass ratio of 0.1--0.2 can drive multiple spirals, and pressure bumps form at locations where the spirals cross the dust ring \citep{Price18}. 
As a result, in addition to dust trapping at the edge of the inner cavity opened by the companion, which forms a dust ring, 
dust grains can also be trapped in the pressure bumps in the dust ring caused by the spirals. 
Consequently, the dust surface density is enhanced in certain azimuthal directions \citep{Price18}. 
Although multiple spirals have been observed in the HD~142527 \citep{Fukagawa06, Casassus12, Avenhaus14, Avenhaus17, Christiaens14, Rodigas14}, 
direct association of the pressure bump with these spirals cannot be made because of the different spatial scales probed by these observations. 
A spiral is expected to induce spiral patterns in a velocity deviation map with a magnitude on the order of 0.1 km s$^{-1}$ \citep[e.g.,][]{Perez18,Pinte19,Rosotti20a}. 
Different from the anticyclonic motion of a vortex, 
close to a spiral in a protoplanetary disk, the gas is radially compressed toward the spiral. 
Such the radial motion close to a spiral (if present) also cannot be detected with the limited resolutions of our data.
Therefore, with the current data sets, we are not able to distinguish a pressure bump formed by a vortex or spirals.
Follow-up observations with higher resolutions are required to resolve the detailed velocity pattern in the dust-trapping region to study the origins of the pressure bump and dust trapping in the HD~142527 disk.

\subsection{Spatial distribution of the maximum grain size}
The spectral index of the dust absorption coefficient $\beta$ can be an assessment of the maximum grain size in a protoplanetary disk \citep{Testi14}.
Our results of the SED fitting of the continuum data at six wavelengths show decreasing $\beta$ from the south to the north in the disk (Fig.~\ref{dust}). 
This could suggest that the maximum grain size is larger in the north and smaller in the south in the HD~142527 disk. 
This trend is consistent with the spatial distribution of the grain size  in the HD~142527 disk inferred from the studies of the dust polarization and scattering \citep{Kataoka16,Ohashi18}. 
Nevertheless, it is uncertain to estimate the maximum grain size from $\beta$ without knowing the compositions of the dust grains or the slope of the grain size distribution \citep[e.g.,][]{Testi14,Birnstiel18}.

Figure~\ref{dv}d compares the velocity deviation from the Keplerian rotation with the $\beta$ distribution in the HD~142527 disk. 
As discussed in Section \ref{trapping}, our results suggest that there is a pressure bump and dust trapping at a radius of 1\farcs2 and position angles of 300\arcdeg--360$\arcdeg$, 
but we do not find any correlation between the $\beta$ distribution and the location where dust trapping occurs. 
This may suggest that there is no significant grain growth in the pressure bump or the region of dust trapping. 
However, we note that if dust scattering is significant, especially in the dust trapping region where the dust density is high, $\beta$ could be overestimated with our analysis (Section \ref{sec_dust}).
If it is the case, the dust $\beta$ in the dust trapping region could be actually lower.

Overall, the dust $\beta$ decreases from the south to north in the HD~142527 disk, suggesting 
more large dust grains in the north than the south. 
This spatial distribution of the grain sizes could be due to asymmetric pressure profiles in the disk. 
Our analysis on the velocity deviation from the Keplerian rotation suggests a stronger pressure gradient in the north than in the south (Fig.~\ref{dv}). 
As shown in the simulations of the interaction between the HD~142527 disk and its central binary system, 
the companion can excite asymmetric spiral patterns in the disk, 
which may induce stronger variations in the northern part of the disk \citep{Price18}. 
As a result, large dust grains in the north may drift inward more slowly compared with those in the south due to the variation in the gas pressure, 
and the distribution of the large dust grains could be more extended in the north. 
Multiple spiral arms have been observed in infrared and the CO lines in the HD~142527 disk \citep{Fukagawa06, Casassus12, Avenhaus14, Avenhaus17, Christiaens14, Rodigas14}.
The spiral arms detected in the CO lines extend toward 4$\arcsec$ away in the north, 
and the northern spiral arms in CO are brighter than those in the south \citep{Casassus12}. 
In the simulations by \citet{Price18},  large dust grains can indeed be more concentrated in the north in the HD~142527 disk by the interaction with the spirals. 

\subsection{Dust feedback in pressure bump}
\begin{figure}
\centering
\includegraphics[width=0.46\textwidth]{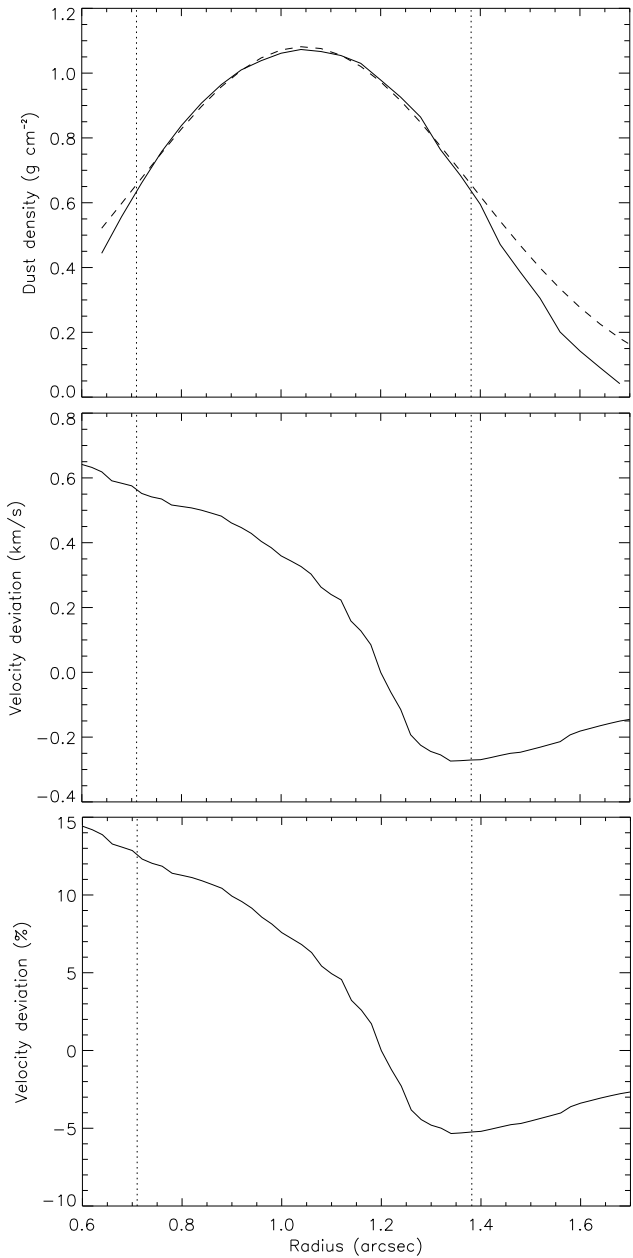}
\caption{Radial profiles of the dust surface density (top panel) and the velocity deviation from the circular Keplerian rotation (middle and bottom panels) in the northwest along the disk major axis, meaning that along a position angle of 340$\arcdeg$. In the top panel, a solid curve presents the observed dust density profile, and a dashed curve shows the fitted Gaussian function to the observed profile. In the middle panel, the velocity deviation is in units of km~s$^{-1}$ after correcting for the inclination by dividing the line-of-sight velocity deviation by $\sin i$. The bottom panel shows the velocity deviation in ratio with the Keplerian velocity, which is computed by dividing the observed velocity deviation map (Fig.~\ref{dv}a) by the best-fit Keplerian velocity map (Fig.~\ref{model}d). Vertical dotted lines denote the locations away from the density peak by the fitted 1$\sigma$ Gaussian width.}\label{dvdr}
\end{figure}

Figure \ref{dvdr} presents radial profiles of the dust surface density and the velocity deviation along the major axis of the disk in the northwest direction.
We fitted a Gaussian function to the radial profile of the dust surface density and measured its 1$\sigma$ width to be 0\farcs34.
The deconvolved width of the dust ring along the major axis can be approximately estimated as $\sqrt{0\farcs34^2-0\farcs19^2}$ to be 0\farcs28, where 0\farcs19 is the 1$\sigma$ width of the beam size, 
and it is comparable to its apparent width.  
The velocity deviation at a radius of 0\farcs6 is $>$10\% of the Keplerian velocity. 
Then it decreases outward and becomes zero at a radius of 1\farcs2. 
Outside the radius of 1\farcs2, the velocity deviation is negative, meaning that the rotational velocity is lower than the Keplerian velocity. 
The minimum velocity deviation is $-$6\% at a radius between 1\farcs3 and 1\farcs4, 
and the difference between the observed rotational velocity and the expected Keplerian velocity becomes smaller at even larger radii.
Therefore, there is a turning point, where the velocity deviation reaches the local maximum in absolute value, in the radial profile of the velocity deviation at a radius between 1\farcs3 and 1\farcs4. 
We note that the radius of the dust density peak is $\sim$0\farcs1 inner than the radius where the velocity deviation is zero, 
and the extension of the dust ring, which is defined as its 1$\sigma$ Gaussian width, approximately reaches the turning point in the radial profile of the velocity deviation.

For a qualitative comparison, we computed the expected velocity deviation due to a Gaussian pressure bump centered at a radius of 1\farcs2 in a Keplerian disk around a central star of 2.48~$M_\odot$ (Fig.~\ref{dpdr}), the same as those in the HD~142527 disk.
We assumed three different radial profiles of the Gaussian pressure bump with different 1$\sigma$ widths (black solid and red dashed curves) or which is on the top of a power-law pressure gradient (blue dashed curve).  
The locations of the 1$\sigma$ width from the center of the pressure bumps are denoted with vertical dotted lines. 
In addition, we computed one more case where the pressure profile at radii larger than 1\farcs2 is a Gaussian function and at radii smaller than 1\farcs2 is a parabolic function to mimic a rapid decrease in pressure at smaller radii. 

For a Gaussian pressure bump (black solid and red dashed curves in Fig.~\ref{dpdr}), 
the magnitude of the velocity deviation approximates to zero toward the two sides away from the pressure bump. 
If there is a power-law pressure gradient in addition to a Gaussian pressure bump (blue dashed curves in Fig.~\ref{dpdr}), 
there remains small negative velocity deviations at outer and inner radii away from the pressure bump.
Our observational results of the HD~142527 disk show an increasing velocity deviation with a decreasing radius at radii inner than the center of the pressure bump. 
This trend is different from our model calculations with the Gaussian pressure profiles (solid and dashed curves in Fig.~\ref{dpdr}), 
but it is similar to the case where the pressure at inner radii decreases as a parabolic function (black dashed-dotted curves in Fig.~\ref{dpdr}).
Thus, this result could suggest that the gas pressure at the inner radii in the HD~142527 disk decreases more rapidly than a Gaussian profile, 
which could be due to the inner cavity carved by the companion in HD~142527.
At radii larger than 1\farcs2, 
the observed profile of the velocity deviation is similar to that in our model calculations,  
where the magnitude of the velocity deviation first increases outward and then decreases toward larger radii and which also shows a turning point in the profile of the velocity deviation.
Furthermore, the observed velocity deviation does not approximate to zero at outer radii close to 1\farcs7.
Thus, this comparison suggests that the pressure profile at radii from 1\farcs2 to 1\farcs7 in the northwest in the HD~142527 could be similar to a Gaussian profile on the top of a power-law pressure gradient (blue dashed curves in Fig.~\ref{dpdr}). 
We note that for a Gaussian pressure profile, its 1$\sigma$ width is always smaller than the separation between the center of the pressure bump and the turning point in the radial profile of the velocity deviation.
In the cases presented in Fig.~\ref{dpdr}, the separations between the centers of the pressure bumps and the turning points are approximately 1.2 to 1.5 times the widths of the pressure bumps.
Our observational results showing a turning point at 0\farcs3 away from the center of the pressure bump in the velocity deviation profile could suggest the width of the pressure bump to be 0\farcs2--0\farcs25.
This width is smaller than or comparable to the width of the dust ring, which is 0\farcs28.

\begin{figure}
\centering
\includegraphics[width=0.46\textwidth]{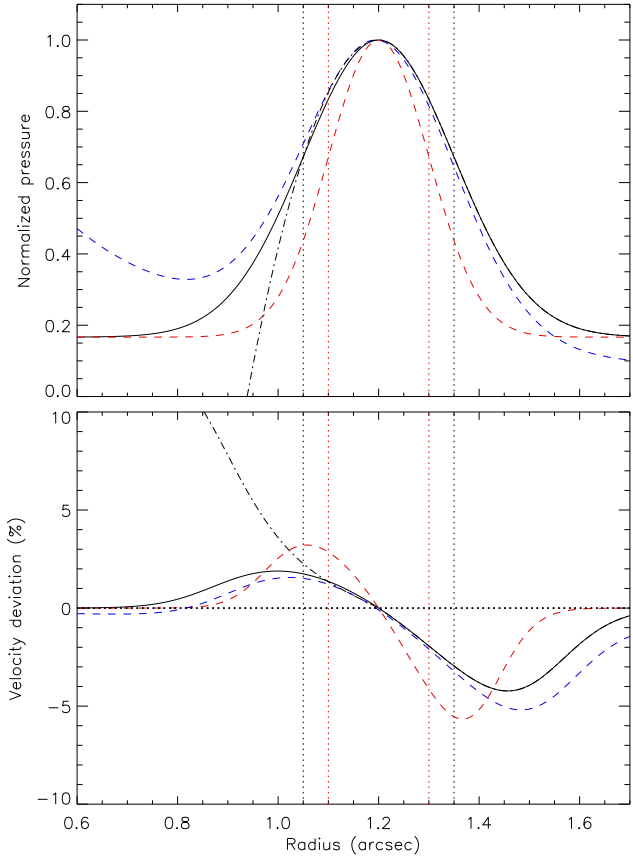}
\caption{Expected velocity deviation from the Keplerian rotation (in ratio with Keplerian velocity; lower panel) due to different radial profiles of gas pressure (upper panel) in a protoplanetary disk. Black solid and red dashed curves present the cases of a Gaussian pressure bump with a 1$\sigma$ width of 0\farcs15 and 0\farcs1 and centered at a radius of 1\farcs2, respectively. Black and red dotted vertical lines denote the locations away from the center of the pressure bump by the 1$\sigma$ Gaussian width. Blue dashed curves present the case of a Gaussian pressure bump with a 1$\sigma$ width of 0\farcs15 and on the top of a power-law pressure gradient. Black dashed-dotted curves present the case where the pressure profile at radii larger than 1\farcs2 is a Gaussian function with a 1$\sigma$ width of 0\farcs15, the same as the black solid curve, but at radii smaller than 1\farcs2 is a parabolic function .
}\label{dpdr}
\label{dpdr}
\end{figure}

Therefore, our results in Fig.~\ref{dvdr} and \ref{dpdr}
could suggest that the 1$\sigma$ width of the dust ring is comparable to or even larger than the width of the pressure bump, in the northwestern part of the disk.
In theoretical calculations of dust trapping by a pressure bump without back reaction of dust onto gas, 
the width of the dust ring is expected to be smaller than the width of the pressure bump, 
when the grain size is large or the turbulence is weak \citep{Dullemond18,Rosotti20}. 
If the grain size is small or the turbulence is strong, 
dust trapping is not effective, 
and the width of the dust ring is comparable to the width of the pressure bump. 
In addition, the gas-to-dust mass ratio is expected to be almost constant in the pressure bump because of the weak dust trapping \citep{Dullemond18}. 
However, our observational results indeed show a lower gas-to-dust mass ratio in the pressure bump than the mean ratio in the disk, suggestive of dust trapping. 
Thus, this scenario of small dust grains or strong turbulence does not match with our observational results, 
and cannot explain the comparable or larger width of the dust ring compared to the width of the pressure bump in the HD~142527 disk.

On the contrary, when dust surface density becomes comparable to gas surface density in a pressure bump in a protoplanetary disk, 
the feedback of the dust on the gas becomes significant \citep{Kanagawa18,Yang20}.
This feedback can alter the gas pressure profile, 
and thus dust radial drift is less efficient in the outer part of the pressure bump. 
Consequently, the dust grain starts to accumulate in the outer part of the pressure bump, 
and the radial extension of the dust ring increases toward outer radii.
With time evolution, the size of a dust ring with the dust feedback can grow to be several times larger than its original size without the dust feedback \citep{Kanagawa18,Yang20}.
Therefore, our observational results, which suggest that the width of the dust ring is comparable to or larger than the width of the pressure bump, could hint at the dust feedback in the northwestern part in the HD~142527 disk.
Nevertheless, further observations at higher angular resolutions to well resolve the radial profiles of the velocity deviation and thus the gas pressure are needed to examine the possible effects of the dust feedback (if indeed present) on the pressure profile in the HD~142527 disk.

\section{Summary}
We analyzed the archival data of the continuum emission at 3~mm, 2.1~mm, 1.3~mm, 0.9~mm, 0.6~mm, and 0.4~mm and $^{13}$CO and C$^{18}$O (1--0, 2--1, and 3--2) lines in the protoplanetary disk around HD~142527 obtained with ALMA.
We performed SED fitting to the continuum data at the six wavelengths with the gray-body slab models and estimated the distributions of the dust temperature, continuum optical depth, and spectral index of dust absorption coefficient $\beta$ in the HD~142527 disk. 
From the estimated continuum optical depth and $\beta$, we inferred the distribution of the dust surface density in the disk. 
To estimate the distribution of the gas surface density, 
we adopted model vertical density and temperature profiles in hydrostatic equilibrium, 
computed the radiative transfer equation on the assumption of the LTE condition, 
integrated the equation along the line of sight, 
and generated model spectra of the C$^{18}$O (3--2), (2--1), and (1--0) lines. 
Then we fitted the model spectra to the observed spectra and estimated gas surface density for every position in the disk.
Finally, with the estimated gas and dust density distributions, we derived the distribution of the gas-to-dust mass ratio.
Our results show that the dust density is a factor of three higher and the gas-to-dust mass ratio is a factor of three lower in the northern part of the HD~142527 disk than the mean values in the disk.
In addition, the dust $\beta$ decreases from two to approximately one from the southern to northern parts of the disk.
These observed trends of the dust distributions and properties are consistent with those reported in the previous studies \citep[e.g.,][]{Casassus15, Muto15, Boehler17, Soon19}.

In addition, 
we performed fitting of the Keplerian disk models to the C$^{18}$O (2--1) and (1--0) data in the visibility domain to measure the disk rotation.
Then we subtracted the peak velocity map of our best-fit disk model from that observed in the C$^{18}$O (2--1) emission to investigate non-Keplerian motion in the HD~142527 disk. 
We found that at the position angles from 300$\arcdeg$ to 360$\arcdeg$, 
the disk rotation is super- and sub-Keplerian inside and outside the radius of the dust ring, 1\farcs2, respectively.
This radial profile of the velocity deviation from Keplerian rotation is consistent with the expectation from the existence of a pressure bump at a radius of 1\farcs2.
However, from the observed velocity patterns, we cannot distinguish whether this pressure bump is formed by a vortex or by spirals driven by the massive companion.
The location of this possible pressure bump is coincident with the region showing the lowest gas-to-dust mass ratio in the HD~142527 disk.
Therefore, our results suggest that the dust grains are trapped by the pressure bump in the northern part of the HD~142527 disk.
On the other hand, although the dust $\beta$ is lower in the north than in the south in the HD~142527 disk, 
suggesting that the maximum grain size is larger in the north, 
there is no spatial correlation between the $\beta$ distribution and the location of the pressure bump. 
Thus, our results may suggest that there is no significant grain growth in the pressure bump.
However, we note that if dust scattering is important in the dust trapping region, where the dust density is high, 
the dust $\beta$ in that region could be actually lower than our current estimates.

We extracted the radial profile of the dust surface density along the disk major axis, which passes through the location of the pressure bump, 
and measured the width of the dust ring. 
We also estimated the upper limit of the width of the pressure bump from the radial profile of the velocity deviation from the Keplerian rotation. 
We found that the width of the dust ring is comparable to or even larger than width of the pressure bump. 
This result cannot be explained with the theoretical models without considering back reaction from dust on gas. 
On the contrary, the comparable or wider width of the dust ring than the pressure bump together with a low gas-to-dust mass ratio is consistent with the expectation from the theoretical models of dust trapping with dust feedback. 
Therefore, our results could hint that dust feedback is important in the northern part of the HD~142527 disk.

\begin{acknowledgements} 
We thank St\'ephane Guilloteau, Ya-Wen Tang, and Bo-Ting Shen for their assistance and advice on fitting the molecular-line data with Keplerian disk models using {\it DiskFit}. 
We thank Shigehisa Takakuwa and Sheng-Yuan Liu  for fruitful suggestions for us to improve our manuscript.
This paper makes use of the following ALMA data: ADS/JAO.ALMA\#2012.1.00631.S, \#2012.1.00725.S, \#2013.1.00305.S, \#2013.1.00670.S, \#2015.1.00614.S, \#2015.1.01137.S, \#2015.1.01353.S, and \#2017.1.00987.S. 
ALMA is a partnership of ESO (representing its member states), NSF (USA) and NINS (Japan), together with NRC (Canada), MOST and ASIAA (Taiwan), and KASI (Republic of Korea), in cooperation with the Republic of Chile. The Joint ALMA Observatory is operated by ESO, AUI/NRAO and NAOJ. 
H.-W.Y. acknowledges support from Ministry of Science and Technology (MOST) in Taiwan through the grant  MOST 108-2112-M-001-003-MY2.
P.-G.G. acknowledges support from MOST in Taiwan through the grant MOST 105-2119-M-001-043-MY3.
\end{acknowledgements}

\begin{appendix}
\section{Continuum and molecular-line images}\label{allim}
\begin{figure*}
\centering
\includegraphics[width=\textwidth]{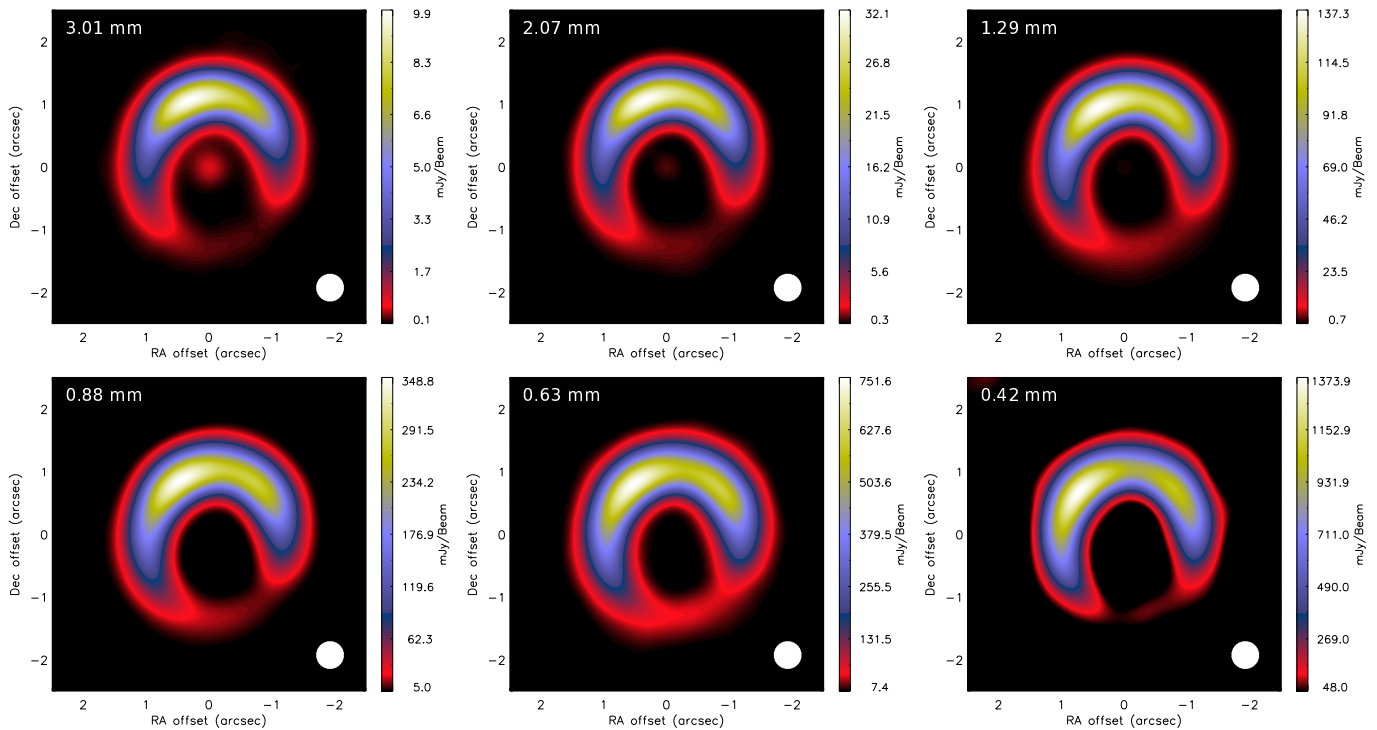}
\caption{Continuum images at six different wavelengths from 3~mm to 0.4~mm obtained with the ALMA observations included in our analysis. The wavelength of each image is labeled  in the upper left corner in each panel. All the images were convolved to a beam size of 0\farcs45, which is shown as white filled circles. Zero offset refers to the stellar position.}
\label{cont_im}
\end{figure*}

\begin{figure*}
\centering
\includegraphics[width=\textwidth]{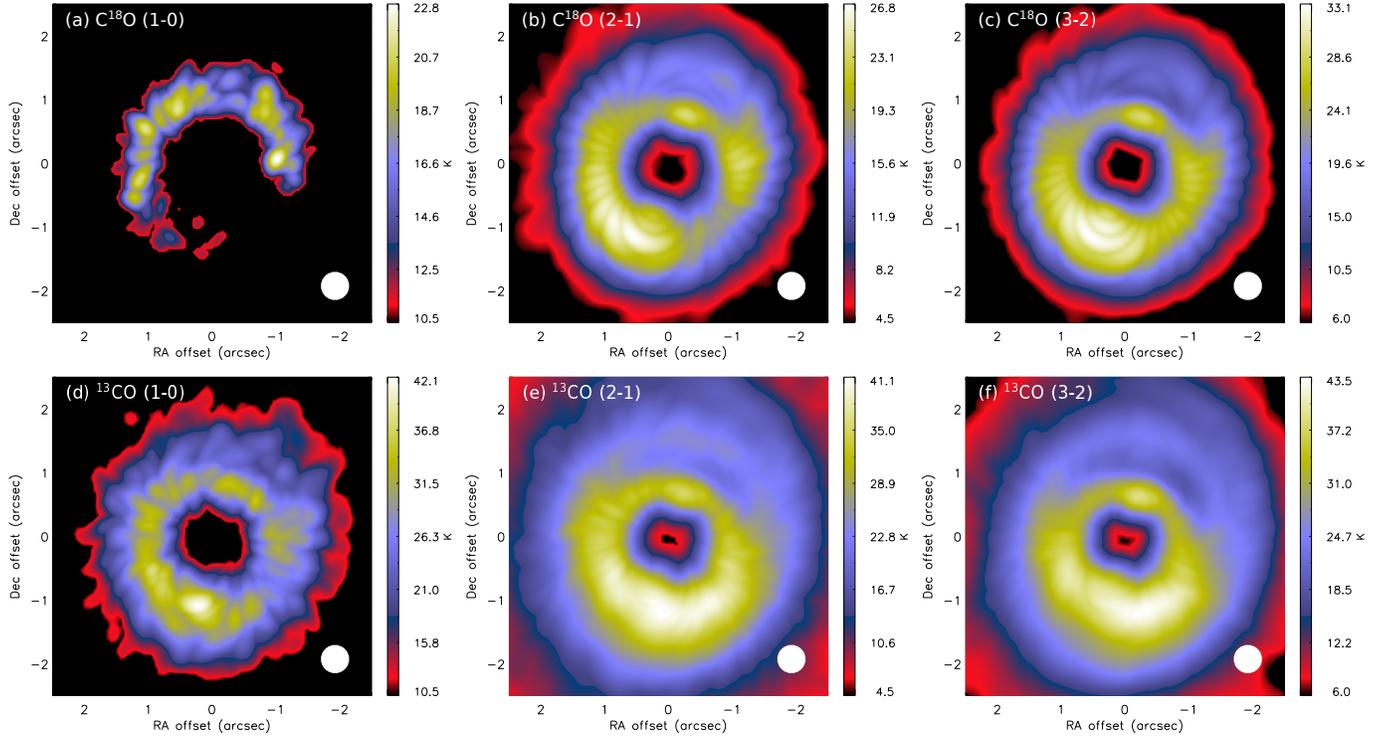}
\caption{Maps of peak brightness temperatures of the $^{13}$CO and C$^{18}$O (1--0), (2--1), and (3--2) emission observed with ALMA included in our analysis. The line and transition of each image are labeled  in the upper left corner in each panel. All the images were convolved to a beam size of 0\farcs45, which is shown as white filled circles. Zero offset refers to the stellar position.}
\label{line_im}
\end{figure*}

Figure \ref{cont_im} and \ref{line_im} present the continuum images at different wavelengths and the peak intensity maps of the $^{13}$CO and C$^{18}$O lines of different transitions, which are analyzed in this work. 
All the images were convolved to a circular beam with a size of 0\farcs45.
The $^{13}$CO and C$^{18}$O peak intensities were converted to brightness temperature. 

\section{Maps of residual velocity due to different errors or mechanisms}\label{simdv}
\begin{figure*}
\centering
\includegraphics[width=\textwidth]{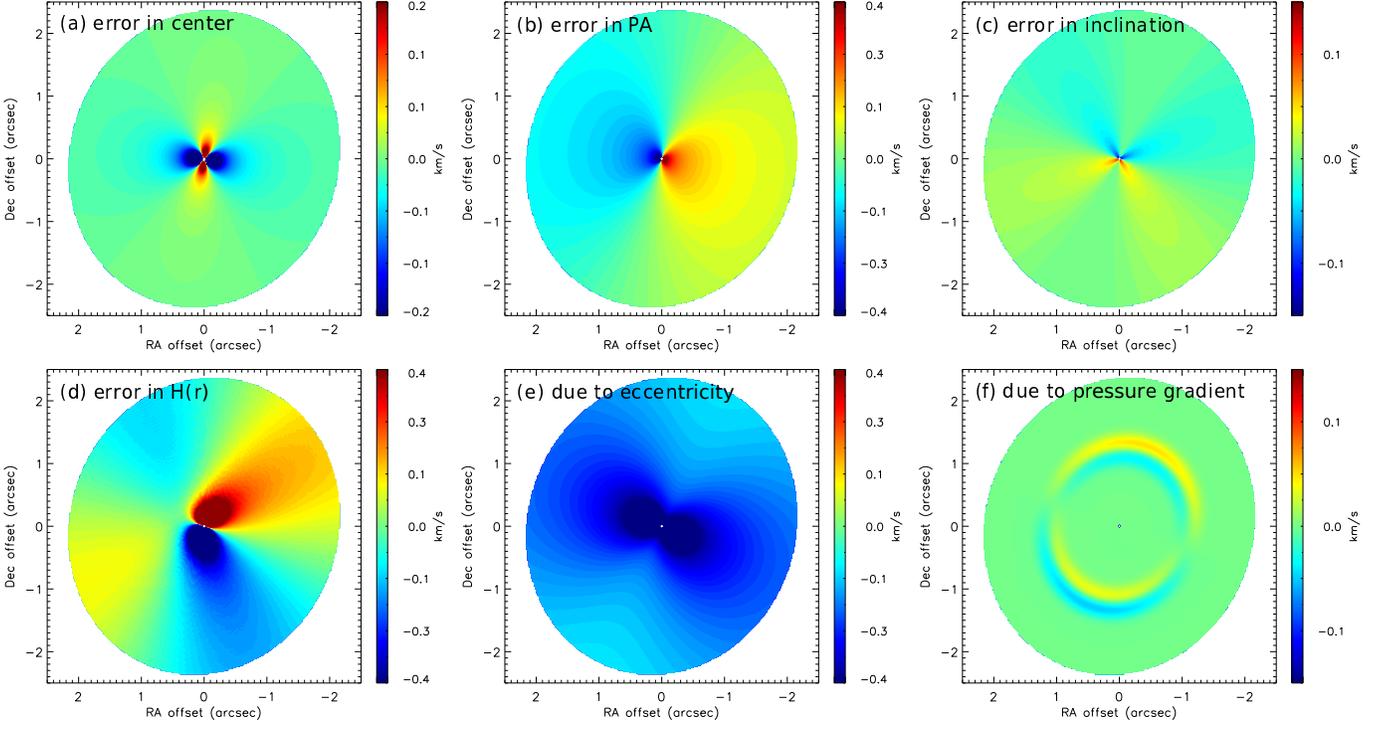}
\caption{Expected residual velocity maps (in units of km\,s$^{-1}$) due to different errors or mechanisms in the disk models. Panel a--d present the cases when there are errors in center, position angle of the major axis, inclination angle, and scale heights of the emitting layer, respectively. Panel e and f present the cases when the models disks have non-zero eccentricity or a axisymmetric Gaussian pressure bump centered at a radius of 1\farcs2. See Appendix \ref{simdv} for details of the calculations.}
\label{dv_im}
\end{figure*}

To demonstrate the patterns of residual velocity caused by different errors in the disk models or by different mechanisms, 
we first computed the expected velocity map of a geometrically thin Keplerian disk around a central star with a mass of 2.48 $M_\odot$. 
The distance was assumed to be 156 pc. 
The position angle of the major axis and inclination angle of the disk were adopted to be 340$\arcdeg$ and 26.7$\arcdeg$. 
These disk parameters were chosen to be the same as those of the HD~142527 disk.
Then we slightly changed the disk parameters and computed new velocity maps to mimic errors in the disk models.
We computed four cases, (a) by shifting the center by 0\farcs04, (b) by changing the position angle by 5$\arcdeg$, (c) by changing the inclination angle by 2$\arcdeg$ and (d) by considering a flaring disk, where the line emission originates from an upper layer but not the midplane in the disk. 
In the case of changing the inclination angle, the stellar mass adopted in the calculations was also changed by 20\% to keep the projected line-of-sight velocity along the major axis unchanged. 
In the case of a flaring disk, the scale height ($h/r$) of the emitting layer was adopted to be $0.15\times(r/1.0)^{1.25}$, where $r$ is the radius in the midplane. 
This $h/r$ profile is adopted to be similar to that of the emitting layer of the $^{13}$CO (2--1) line in the protoplanetary disk around HD~163296 \citep{Teague18}.
Finally, we subtracted the original velocity map from these velocity maps with the given errors, 
and their corresponding residual velocity maps are shown in Fig.~\ref{dv_im}.
Thus, these maps show the resultant patterns of the residual velocity, if the center, position angle, inclination angle, and scale height are inaccurate in the disk models.

In addition to the errors in the disk models, we also computed two cases assuming that the disk has non-zero eccentricity and a pressure bump. 
In the case of an eccentric disk, a constant eccentricity of 0.2 was adopted in the entire disk, and the position angle of the pericenter was adopted to be 330$\arcdeg$. 
These values are similar to those measured from the dust ring traced by the continuum emission \citep{Boehler17}.
We followed the equations in \citet{Goodchild06} to compute the motional velocity of an eccentric disk.
The residual velocity of an eccentric disk after subtracting the circular Keplerian rotation is all blueshifted. 
That is because the rotational velocity is faster and slower than the circular Keplerian velocity at the pericenter and apocenter, respectively. 
In addition, there are radial velocities toward and away from the stellar position when the gas is moving from the apocenter to the pericenter and from the pericenter to the apocenter, respectively \citep[e.g.,][]{Hsieh12}.
In the case of the presence of a pressure bump in the disk, 
we assumed that the radial profile of the pressure bump is a Gaussian function with its center at 1\farcs2, and the pressure profile is axisymmetric in the disk.
The rotational velocity was computed from the gravitational force and the pressure gradient, following the equations in \citet{Rosotti20}.
Because of the pressure gradient, the rotational velocity is super- and sub-Keplerian inside and outside the center of the pressure bump. 
The residual velocity maps, after subtracting the circular Keplerian rotation from the velocity maps of these two cases, are shown in Fig.~\ref{dv_im}e and f, 
and these maps present the patterns of the residual velocity if the disk is eccentric or has a pressure bump.

\end{appendix}

\software{CASA \citep{McMullin07}; DiskFit \citep{Pietu07}; GILDAS \citep{Pety05,gildas13}}

\begin{thebibliography}{}
\bibitem[Ataiee et al.(2013)]{Ataiee13} Ataiee, S., Pinilla, P., Zsom, A., et al.\ 2013, \aap, 553, L3
\bibitem[Avenhaus et al.(2017)]{Avenhaus17} Avenhaus, H., Quanz, S.~P., Schmid, H.~M., et al.\ 2017, \aj, 154, 33
\bibitem[Avenhaus et al.(2014)]{Avenhaus14} Avenhaus, H., Quanz, S.~P., Schmid, H.~M., et al.\ 2014, \apj, 781, 87
\bibitem[Baruteau et al.(2019)]{Baruteau19} Baruteau, C., Barraza, M., P{\'e}rez, S., et al.\ 2019, \mnras, 486, 304
\bibitem[Baruteau \& Zhu(2016)]{Baruteau16} Baruteau, C. \& Zhu, Z.\ 2016, \mnras, 458, 3927
\bibitem[Beckwith et al.(1990)]{Beckwith90} Beckwith, S.~V.~W., Sargent, A.~I., Chini, R.~S., et al.\ 1990, \aj, 99, 924
\bibitem[Biller et al.(2012)]{Biller12} Biller, B., Lacour, S., Juh{\'a}sz, A., et al.\ 2012, \apjl, 753, L38
\bibitem[Birnstiel et al.(2010)]{Birnstiel10} Birnstiel, T., Dullemond, C.~P., \& Brauer, F.\ 2010, \aap, 513, A79
\bibitem[Birnstiel et al.(2018)]{Birnstiel18} Birnstiel, T., Dullemond, C.~P., Zhu, Z., et al.\ 2018, \apjl, 869, L45
\bibitem[Boehler et al.(2017)]{Boehler17} Boehler, Y., Weaver, E., Isella, A., et al.\ 2017, \apj, 840, 60
\bibitem[Casassus et al.(2015a)]{Casassus15a} Casassus, S., Marino, S., P{\'e}rez, S., et al.\ 2015, \apj, 811, 92
\bibitem[Casassus et al.(2012)]{Casassus12} Casassus, S., Perez M., S., Jord{\'a}n, A., et al.\ 2012, \apjl, 754, L31
\bibitem[Casassus et al.(2015b)]{Casassus15} Casassus, S., Wright, C.~M., Marino, S., et al.\ 2015, \apj, 812, 126
\bibitem[Cazzoletti et al.(2018)]{Cazzoletti18} Cazzoletti, P., van Dishoeck, E.~F., Pinilla, P., et al.\ 2018, \aap, 619, A161
\bibitem[Casassus et al.(2013)]{Casassus13} Casassus, S., van der Plas, G.~M., Perez, S., et al.\ 2013, \nat, 493, 191
\bibitem[Chiang \& Youdin(2010)]{Chiang10} Chiang, E. \& Youdin, A.~N.\ 2010, Annual Review of Earth and Planetary Sciences, 38, 493
\bibitem[Christiaens et al.(2018)]{Christiaens18} Christiaens, V., Casassus, S., Absil, O., et al.\ 2018, \aap, 617, A37
\bibitem[Christiaens et al.(2014)]{Christiaens14} Christiaens, V., Casassus, S., Perez, S., et al.\ 2014, \apjl, 785, L12
\bibitem[Dartois et al.(2003)]{Dartois03} Dartois, E., Dutrey, A., \& Guilloteau, S.\ 2003, \aap, 399, 773
\bibitem[Dullemond et al.(2018)]{Dullemond18} Dullemond, C.~P., Birnstiel, T., Huang, J., et al.\ 2018, \apjl, 869, L46
\bibitem[Fedele et al.(2017)]{Fedele17} Fedele, D., Carney, M., Hogerheijde, M.~R., et al.\ 2017, \aap, 600, A72
\bibitem[Francis \& van der Marel(2020)]{Francis20} Francis, L., \& van der Marel, N.\ 2020, \apj, 892, 111
\bibitem[Fukagawa et al.(2006)]{Fukagawa06} Fukagawa, M., Tamura, M., Itoh, Y., et al.\ 2006, \apjl, 636, L153
\bibitem[Fukagawa et al.(2013)]{Fukagawa13} Fukagawa, M., Tsukagoshi, T., Momose, M., et al.\ 2013, \pasj, 65, L14
\bibitem[Furuya \& Aikawa(2014)]{Furuya14} Furuya, K., \& Aikawa, Y.\ 2014, \apj, 790, 97
\bibitem[Gaia Collaboration et al.(2018)]{Gaia18} Gaia Collaboration, Brown, A.~G.~A., Vallenari, A., et al.\ 2018, \aap, 616, A1
\bibitem[Gaia Collaboration et al.(2016)]{Gaia16} Gaia Collaboration, Prusti, T., de Bruijne, J.~H.~J., et al.\ 2016, \aap, 595, A1
\bibitem[Gildas Team(2013)]{gildas13} Gildas Team\ 2013, Astrophysics Source Code Library. ascl:1305.010
\bibitem[Goodchild \& Ogilvie(2006)]{Goodchild06} Goodchild, S. \& Ogilvie, G.\ 2006, \mnras, 368, 1123
\bibitem[Lee \& Gu(2015)]{Lee15} Lee, W.-K. \& Gu, P.-G.\ 2015, \apj, 814, 72
\bibitem[Helled et al.(2014)]{Helled14} Helled, R., Bodenheimer, P., Podolak, M., et al.\ 2014, Protostars and Planets VI, 643
\bibitem[Hsieh \& Gu(2012)]{Hsieh12} Hsieh, H.-F. \& Gu, P.-G.\ 2012, \apj, 760, 119
\bibitem[Huang et al.(2018)]{Huang18} Huang, P., Isella, A., Li, H., et al.\ 2018, \apj, 867, 3
\bibitem[J{\o}rgensen et al.(2004)]{Jorgensen04} J{\o}rgensen, J.~K., Sch{\"o}ier, F.~L., \& van Dishoeck, E.~F.\ 2004, \aap, 416, 603
\bibitem[Kanagawa et al.(2018)]{Kanagawa18} Kanagawa, K.~D., Muto, T., Okuzumi, S., et al.\ 2018, \apj, 868, 48
\bibitem[Kataoka et al.(2016)]{Kataoka16} Kataoka, A., Tsukagoshi, T., Momose, M., et al.\ 2016, \apjl, 831, L12
\bibitem[Lacour et al.(2016)]{Lacour16} Lacour, S., Biller, B., Cheetham, A., et al.\ 2016, \aap, 590, A90
\bibitem[Lin(2014)]{Lin14} Lin, M.-K.\ 2014, \mnras, 437, 575
\bibitem[Liu(2019)]{Liu19} Liu, H.~B.\ 2019, \apjl, 877, L22
\bibitem[Mangum \& Shirley(2015)]{Mangum15} Mangum, J.~G., \& Shirley, Y.~L.\ 2015, \pasp, 127, 266
\bibitem[McClure et al.(2016)]{McClure16} McClure, M.~K., Bergin, E.~A., Cleeves, L.~I., et al.\ 2016, \apj, 831, 167
\bibitem[McMullin et al.(2007)]{McMullin07} McMullin, J. P., Waters, B., Schiebel, D., et al.\ 2007, Astronomical Data Analysis Software and Systems XVI (ASP Conf. Ser. 376), ed. R. A. Shaw, F. Hill, \& D. J. Bell (San Francisco, CA: ASP), 127
\bibitem[Miotello et al.(2016)]{Miotello16} Miotello, A., van Dishoeck, E.~F., Kama, M., et al.\ 2016, \aap, 594, A85
\bibitem[Muto et al.(2015)]{Muto15} Muto, T., Tsukagoshi, T., Momose, M., et al.\ 2015, \pasj, 67, 122
\bibitem[Ohashi(2008)]{Ohashi08} Ohashi, N.\ 2008, \apss, 313, 101
\bibitem[Ohashi et al.(2018)]{Ohashi18} Ohashi, S., Kataoka, A., Nagai, H., et al.\ 2018, \apj, 864, 81
\bibitem[P{\'e}rez et al.(2014)]{Perez14} P{\'e}rez, L.~M., Isella, A., Carpenter, J.~M., et al.\ 2014, \apjl, 783, L13
\bibitem[Perez et al.(2015)]{Perez15} Perez, S., Casassus, S., M{\'e}nard, F., et al.\ 2015, \apj, 798, 85
\bibitem[P{\'e}rez et al.(2018)]{Perez18} P{\'e}rez, S., Casassus, S., \& Ben{\'\i}tez-Llambay, P.\ 2018, \mnras, 480, L12
\bibitem[Pety(2005)]{Pety05} Pety, J.\ 2005, SF2A-2005: Semaine de l'Astrophysique Francaise, 721
\bibitem[Pi{\'e}tu et al.(2007)]{Pietu07} Pi{\'e}tu, V., Dutrey, A., \& Guilloteau, S.\ 2007, \aap, 467, 163
\bibitem[Pinilla et al.(2017)]{Pinilla17} Pinilla, P., P{\'e}rez, L.~M., Andrews, S., et al.\ 2017, \apj, 839, 99
\bibitem[Pinilla et al.(2012)]{Pinilla12} Pinilla, P., Benisty, M., \& Birnstiel, T.\ 2012, \aap, 545, A81
\bibitem[Pinte et al.(2018)]{Pinte18} Pinte, C., M{\'e}nard, F., Duch{\^e}ne, G., et al.\ 2018, \aap, 609, A47
\bibitem[Pinte et al.(2019)]{Pinte19} Pinte, C., van der Plas, G., M{\'e}nard, F., et al.\ 2019, Nature Astronomy, 3, 1109
\bibitem[Price et al.(2018)]{Price18} Price, D.~J., Cuello, N., Pinte, C., et al.\ 2018, \mnras, 477, 1270
\bibitem[Ragusa et al.(2017)]{Ragusa17} Ragusa, E., Dipierro, G., Lodato, G., et al.\ 2017, \mnras, 464, 1449
\bibitem[Raymond et al.(2014)]{Raymond14} Raymond, S.~N., Kokubo, E., Morbidelli, A., et al.\ 2014, Protostars and Planets VI, 595
\bibitem[Robert et al.(2020)]{Robert20} Robert, C.~M.~T., M{\'e}heut, H., \& M{\'e}nard, F.\ 2020, \aap, 641, A128
\bibitem[Rodigas et al.(2014)]{Rodigas14} Rodigas, T.~J., Follette, K.~B., Weinberger, A., et al.\ 2014, \apjl, 791, L37
\bibitem[Rosenfeld et al.(2013)]{Rosenfeld13} Rosenfeld, K.~A., Andrews, S.~M., Hughes, A.~M., et al.\ 2013, \apj, 774, 16
\bibitem[Rosotti et al.(2020a)]{Rosotti20a} Rosotti, G.~P., Benisty, M., Juh{\'a}sz, A., et al.\ 2020, \mnras, 491, 1335
\bibitem[Rosotti et al.(2020b)]{Rosotti20} Rosotti, G.~P., Teague, R., Dullemond, C., et al.\ 2020, \mnras, 495, 173
\bibitem[Schwarz et al.(2016)]{Schwarz16} Schwarz, K.~R., Bergin, E.~A., Cleeves, L.~I., et al.\ 2016, \apj, 823, 91
\bibitem[Soon et al.(2019)]{Soon19} Soon, K.-L., Momose, M., Muto, T., et al.\ 2019, \pasj, 71, 124
\bibitem[Teague et al.(2018)]{Teague18} Teague, R., Bae, J., Bergin, E.~A., et al.\ 2018, \apjl, 860, L12
\bibitem[Testi et al.(2014)]{Testi14} Testi, L., Birnstiel, T., Ricci, L., et al.\ 2014, Protostars and Planets VI, 339
\bibitem[Ueda et al.(2020)]{Ueda20} Ueda, T., Kataoka, A., \& Tsukagoshi, T.\ 2020, \apj, 893, 125
\bibitem[van der Marel et al.(2016a)]{Marel16} van der Marel, N., Cazzoletti, P., Pinilla, P., et al.\ 2016, \apj, 832, 178
\bibitem[van der Marel et al.(2016b)]{Marel16b} van der Marel, N., van Dishoeck, E.~F., Bruderer, S., et al.\ 2016, \aap, 585, A58
\bibitem[van der Marel et al.(2015)]{Marel15} van der Marel, N., Pinilla, P., Tobin, J., et al.\ 2015, \apjl, 810, L7
\bibitem[Williams \& Cieza(2011)]{Williams11} Williams, J.~P. \& Cieza, L.~A.\ 2011, \araa, 49, 67
\bibitem[Wilson \& Rood(1994)]{Wilson94} Wilson, T.~L. \& Rood, R.\ 1994, \araa, 32, 191
\bibitem[Yamaguchi et al.(2020)]{Yamaguchi20} Yamaguchi, M., Akiyama, K., Tsukagoshi, T., et al.\ 2020, \apj, 895, 84
\bibitem[Yang \& Zhu(2020)]{Yang20} Yang, C.-C. \& Zhu, Z.\ 2020, \mnras, 491, 4702
\bibitem[Yen et al.(2019)]{Yen19} Yen, H.-W., Gu, P.-G., Hirano, N., et al.\ 2019, \apj, 880, 69
\bibitem[Zhang et al.(2018)]{Zhang18} Zhang, S., Zhu, Z., Huang, J., et al.\ 2018, \apjl, 869, L47
\bibitem[Zhu et al.(2014)]{Zhu14} Zhu, Z., Stone, J.~M., Rafikov, R.~R., et al.\ 2014, \apj, 785, 122
\bibitem[Zhu et al.(2019)]{Zhu19} Zhu, Z., Zhang, S., Jiang, Y.-F., et al.\ 2019, \apjl, 877, L18
\end{thebibliography}
\end{document}